\documentstyle[prl,aps,multicol,epsfig,epsf,psfig,array]{revtex}
\newcommand{\beq}{\begin{equation}}
\newcommand{\eeq}{\end{equation}}
\newcommand{\bqa}{\begin{eqnarray}}
\newcommand{\eqa}{\end{eqnarray}}

\begin{document} 
\setlength{\baselineskip}{0.333333in}
\tightenlines

\draft
\author{U. Al Khawaja$\;^1$, J.O. Andersen$\;^1$,
N.P. Proukakis$\;^{1,2}$, 
and H.T.C Stoof$\;^1$
}
\address{\it $^1$Institute for Theoretical Physics, 
\\Utrecht University,
Leuvenlaan 4, 3584 CE Utrecht, The Netherlands
\\$^{2}$Foundation for Research and Technology Hellas, \\
Institute of Electronic Structure and Laser, P.O. Box 1527, Heraklion 71 110, Crete, Greece
\\ (\today)}

\title{Low-dimensional Bose gases}

\maketitle

\begin{abstract}
We present an improved many-body T-matrix theory for partially Bose-Einstein
condensed atomic gases by treating the phase fluctuations
exactly. The resulting mean-field theory is valid in arbitrary
dimensions and able to describe the low-temperature crossover
between three, two and one-dimensional Bose gases.
When applied to a degenerate two-dimensional atomic hydrogen gas, 
we obtain a 
reduction of the three-body recombination rate 
which compares favorably with experiment. 
Supplementing the mean-field theory with a renormalization-group approach 
to treat the critical fluctuations, we also incorporate into the theory 
the Kosterlitz-Thouless transition that occurs in a homogeneous
Bose gas in two dimensions. 
In particular, we
calculate the critical conditions for the Kosterlitz-Thouless phase transition
as a function of the microscopic parameters of the theory.
The proposed theory is further applied to a
trapped one-dimensional Bose gas, where we find good
agreement with exact numerical results obtained by
solving a nonlinear Langevin field equation. 

\end{abstract}


\begin{multicols}{2}

\section{Introduction}
Low-dimensional Bose gases have recently attracted attention 
both experimentally and theoretically. 
The interest in these systems stems from the fact that the physics
of low-dimensional systems is fundamentally different from the
physics of systems in three dimensions.
One and two-dimensional Bose-Einstein condensates have recently
been created in the experiment 
of G\"orlitz {\it et al.}~\cite{lowdketterle}. 
This was achieved by lowering the mean-field interaction energy 
in a three-dimensional condensate below the 
energy splitting of either one or two of the directions of the harmonic trap,  
to obtain a two-dimensional or one-dimensional condensate, respectively.  
In a number of other experiments a one-dimensional Bose-Einstein condensate 
was also created in a $^6$Li-$^7$Li mixture~\cite{exp100} and on 
a microchip~\cite{{exp200},{exp300}}.  

Theoretically, low-dimensional Bose gases are particularly interesting due 
to the enhanced importance of phase fluctuations~\cite{mullin,jason,2d,1d}. 
Because of these fluctuations, Bose-Einstein condensation 
cannot take place in a homogeneous one-dimensional 
Bose gas at all temperatures and in a homogeneous two-dimensional Bose gas at 
any nonzero temperature. This is formalized in the Mermin-Wagner-Hohenberg 
theorem~\cite{mermin,hohen}. 
Since this theorem is valid only in the thermodynamic limit, it does not
apply to trapped Bose gases. Therefore, the question 
arises whether under certain conditions we are dealing with  
a true condensate, where the phase is coherent over 
a distance of the order of the size of the system, 
or only with a so-called ``quasicondensate''~\cite{popov},
where the phase 
is coherent over a distance less than the size of the 
system~\cite{mullin,jason,2d,1d}. 
This is one of the main questions, which we address 
quantitatively in this paper. 

In the successful Popov theory for three-dimensional partially Bose-Einstein 
condensed gases, the phase fluctuations are taken into account 
up to the second order around the mean field. 
In view of the above-mentioned importance of phase fluctuations 
in lower dimensions, this is 
insufficient in general and leads to infrared divergences. 
In previous work by three of us, the 
phase fluctuations were taken into account exactly~\cite{jh}. 
The result is a mean-field theory which is free of the 
infrared divergences in all dimensions.
In the present paper, we first review this modified Popov theory 
and then extend it to the many-body T-matrix theory, 
by including the effect of the medium on the scattering properties 
of the atoms in the gas. The present approach improves on previous
attempts by Petrov {\it et al.}~\cite{2d,1d} to describe low-dimensional 
Bose gases by explicitly incorporating also the effect of density 
fluctuations into the theory. As a result  both quantum and thermal
depletion of the
(quasi)condensate can now be accounted for and the theory is no longer only 
valid at very low temperatures where the depletion, and therefore the 
thermal component in the gas, is negligible. 
Most importantly, we present an equation of state for the 
low-dimensional Bose gas that 
is free of infrared divergences and thus valid in any dimension. 
For a trapped Bose gase this implies that we can determine, for a given 
number of atoms, the 
density profile of both the (quasi)condensate and the thermal cloud in 
the gas for any aspect ratio of the trap. 
Also the interesting crossover problem from a 
three-dimensional Bose gas to a one or two-dimensional 
one, that is presently being explored exprimentally \cite{lowdketterle}, 
can be addressed as well.

In the present paper we first use the modified many-body T-matrix theory to 
calculate the one-particle density matrix and determine its 
off-diagonal long-range order. We also calculate the fractional 
depletion of the (quasi)condensate at zero temperature in 
one, two, and three dimensions.
Next, we study the two-dimensional homogeneous Bose gas in considerable detail.
After having 
included the phase fluctuations due to vortex pairs 
by a renormalization group approach, 
we apply the modified many-body T-matrix theory 
to perform an {\it ab initio} study of the
Kosterlitz-Thouless phase transition~\cite{koster}  
from the superfluid to the normal state. 
Since this is a topological phase transition, it cannot be described within
mean-field theory.
Therefore, we proceed as follows. We first use the modified many-body 
T-matrix theory to calculate the quasicondensate density 
and the fugacity of vortices. These results are then used as 
initial conditions 
for a Kosterlitz renormalization group calculation. 
In this manner, we are incorporating critical fluctuations and 
able to calculate nonuniversal
quantities such as the critical temperature for
the Kosterlitz-Thouless phase transition as a function of the density and 
the microscopic parameters of the theory.

Finally, we apply the theory to a trapped one-dimensional
Bose gas, where we calculate the density profile at different temperatures. 
From this we extract the crossover temperature 
for the appearance of a (quasi)condensate as a 
function of the interaction strength. 
We also calculate the behavior of the phase correlation function 
which determines whether there exists a true condensate or only a 
quasicondensate. These predictions are 
compared to exact results based on a stochastic nonlinear field 
equation for the Bose gas~\cite{Stoof_Booklet}.

The paper is organized as follows.
In Sec.~II, we present and discuss the Popov theory and 
its infrared problems. We also present our modified mean-field theory in the
homogeneous limit. 
In Sec.~III, we compare
the latter with exact results in one dimension, and with results
obtained in the Popov approximation in two and three dimensions.
We also calculate the reduction of the three-body recombination rate
for a two-dimensional hydrogen gas and compare it to the experiment 
of Safonov {\it et al.}~\cite{saf}.
In Sec.~IV, we study the Kosterlitz-Thouless phase transition
and in Sec.~V, we generalize the many-body T-matrix theory to inhomogeneous 
situations. In particular, we consider then a one-dimensional trapped Bose 
gase, for which we compare our predictions to numerically exact results.
Finally, we conclude and summarize in Sec.~VI.

\section{Modified Popov Theory}
\label{usapopov}
In this section, we derive the modified Popov theory by treating
the phase fluctuations exactly. 
We also discuss how to incorporate many-body effects into
the theory.
Finally, we give additional 
arguments for the correctness of our approach by using 
an effective action for the density and phase dynamics in a superfluid 
system that is known 
to give exact results in the long-wavelength limit.
\subsection{Phase fluctuations}
\label{phsec}
In order to explain the infrared problems associated with the
phase fluctuations of the condensate most clearly, we first treat a homogeneous
Bose gas in a box of volume $V$. 
Later we generalize to the inhomogeneous case.
The starting point is the grand-canonical Hamiltonian in second-quantized 
language
\bqa\nonumber
{ H}&=&
\int\;d{\bf x}\;
{\hat\psi}^{\dagger}({\bf x})\Bigg[
-{\hbar^2\over2m}\nabla^2-\mu
\Bigg]{\hat\psi}({\bf x})
\\&& \nonumber
+{1\over2}\int\;d{\bf x}\int\;d{\bf x}^{\prime}
{\hat\psi}^{\dagger}({\bf x}){\hat\psi}^{\dagger}({\bf x}^{\prime})
\\&&
\times
V({\bf x}-{\bf x}^{\prime})
{\hat\psi}({\bf x^{\prime}}){\hat\psi}({\bf x})\;,
\eqa
where $\mu$ is the chemical potential, 
and 
$V({\bf x})$ is the atomic two-body interaction potential.
The mass of the atoms is denoted by $m$, and ${\hat\psi}^{\dagger}({\bf x})$
and ${\hat\psi}({\bf x})$ are the usual creation and annihilation field 
operators, respectively.

In the Bose systems considered here and those realized in experiment, the
temperatures are so low that only $s$-wave scattering is important.
Consequently, it is convenient to neglect the momentum dependence 
of the interatomic interaction and 
use $V({\bf x}-{\bf x}^{\prime})=V_0\delta({\bf x}-{\bf x}^{\prime})$. 
In principle this leads to ultraviolet divergences, 
but these can easily be dealt with, as we show later on. 
The Hamiltonian then reduces to
\bqa\nonumber
{H}&=&
\int\;d{\bf x}\;
{\hat\psi}^{\dagger}({\bf x})\Bigg[
-{\hbar^2\over2m}\nabla^2-\mu
\Bigg]{\hat\psi}({\bf x})
\\&& 
+{1\over2}\int\;d{\bf x}V_0
{\hat\psi}^{\dagger}({\bf x}){\hat\psi}^{\dagger}({\bf x})
{\hat\psi}({\bf x}){\hat\psi}({\bf x})\;.
\label{ham}
\eqa

In the presence of a Bose-Einstein condensate 
the annihilation operator is parametrized as
\bqa
\label{paraxy}
\hat{\psi}({\bf x})=\sqrt{n_0}+\hat{\psi}^{\prime}({\bf x})\;,
\eqa
where $n_0$ is the condensate density and $\hat{\psi}^{\prime}({\bf x})$
describes the fluctuations. The standard one-loop expressions for the density
$n$ and the chemical potential $\mu$ are obtained after 
a quadratic approximation to the Hamiltonian in Eq.~(\ref{ham}), 
i.e., by neglecting terms that are of third and fourth order in the 
fluctuations. This yields \cite{{popov},{henk1}}
\bqa\nonumber
n&=&n_0+{1\over V}\sum_{\bf k}
\Bigg[
{\epsilon_{\bf k}+n_0V_0-\hbar\omega_{\bf k}\over2\hbar\omega_{\bf k}}
\\
&&\hspace{0.45cm}+
{\epsilon_{\bf k}+n_0V_0\over\hbar\omega_{\bf k}}
N(\hbar\omega_{\bf k})
\label{1ln}
\Bigg]\;, \\ \nonumber
{\mu\over V_0}&=&n_0+{1\over V}
\sum_{\bf k}
\Bigg[
{2\epsilon_{\bf k}+n_0V_0-2\hbar\omega_{\bf k}\over2\hbar\omega_{\bf k}}
\\
&&
\hspace{0.85cm}+
{2\epsilon_{\bf k}+n_0V_0\over\hbar\omega_{\bf k}}
N(\hbar\omega_{\bf k})\Bigg]\;,
\label{1lmu}
\eqa
where 
$\hbar\omega_{\bf k}=(\epsilon_{\bf k}^2+2n_0V_0\epsilon_{\bf k})^{1/2}$
is the Bogoliubov dispersion relation, $N(x)=1/(e^{\beta x}-1)$ is the
Bose-Einstein distribution function, and $\beta=1/k_{\rm B}T$
is the inverse thermal energy. 

In agreement with the Mermin-Wagner-Hohenberg theorem, the
momentum sums in Eqs.~(\ref{1ln}) and (\ref{1lmu}) contain terms
that are infrared divergent at all temperatures in one dimension
and at any nonzero
temperature in two dimensions. The physical
reason for these ``dangerous'' terms is that the above expressions
have been derived by taking into account only quadratic
fluctuations around the classical result $n_0$, i.e., by writing
the annihilation operator for the atoms as $\hat{\psi}({\bf x}) =
\sqrt{n_0} + \hat{\psi}'({\bf x})$ and neglecting in the
Hamiltonian terms of third and fourth order in $\hat{\psi}'({\bf
x})$. As a result the phase fluctuations of the condensate give
the quadratic contribution $n_0 \langle \hat{\chi}({\bf x})
\hat{\chi}({\bf x}) \rangle$ to the right-hand side of the above
equations, whereas an exact approach that sums up all the
higher-order terms in the expansion would clearly give no
contribution at all to these local quantities because $n_0\langle
e^{-i\hat{\chi}({\bf x})} e^{i\hat{\chi}({\bf x})} \rangle = n_0\left(1 +
\langle \hat{\chi}({\bf x})\hat{\chi}({\bf x}) \rangle + \dots\right) =1$. 
To correct for this we thus need to subtract the quadratic
contribution of the phase fluctuations, which from
Eqs.~(\ref{1ln}) and (\ref{1lmu}) is seen to be given by
\begin{equation}
\label{phase}
n_0 \langle \hat{\chi}({\bf x}) \hat{\chi}({\bf x}) \rangle =
{1\over V} \sum_{\bf k} {n_0 V_0\over2\hbar\omega_{\bf k}}
\left[1+2N(\hbar\omega_{\bf k})\right]\;.
\end{equation}
As expected, the infrared divergences that occur in the one and
two-dimensional cases are removed by performing this subtraction.

After having removed the spurious contributions from the phase
fluctuations of the condensate, the resulting expressions turn out
to be ultraviolet divergent. These divergences are removed by
the standard renormalization of the bare coupling constant $V_0$.
Apart from a subtraction, this essentially amounts to replacing
everywhere the bare two-body potential $V_0$ by the two-body
T-matrix evaluated at zero initial and final relative momenta and
at the energy $-2\mu$, which we denote from now on by $T^{\rm
2B}(-2\mu)$. It is formally defined by
\bqa
{1\over T^{\rm 2B}(-2\mu)}&=&{1\over V_0}+{1\over V}\sum_{\bf k}
{1\over2\epsilon_{\bf k}+2\mu}\;.
\eqa
Note that the energy argument of the T-matrix is
$-2\mu$, because this is precisely the energy it costs to excite two
atoms from the condensate \cite{schick,fisher}. 
After renormalization, the density and chemical potential are
\bqa\nonumber
n&=&n_0+{1\over V}\sum_{\bf k}
\Bigg[
{\epsilon_{\bf k}-\hbar\omega_{\bf k}\over2\hbar\omega_{\bf k}}
+{n_0T^{\rm 2B}(-2\mu)\over2\epsilon_{\bf k}+2\mu} 
\\&&+
{\epsilon_{\bf k}\over\hbar\omega_{\bf k}}
N(\hbar\omega_{\bf k})
\Bigg]\;, 
\label{h1}
\\ \nonumber
\mu&=&(2n-n_0)T^{\rm 2B}(-2\mu)  \\
\label{h2} 
&=&(2n'+n_0)T^{\rm 2B}(-2\mu) \;,
\eqa
where $n'=n-n_0$ represents the depletion of the condensate
due to quantum and thermal fluctuations and
the Bogoliubov quasiparticle dispersion now equals
$\hbar\omega_{\bf k}
=\left[\epsilon_{\bf k}^2+2n_0T^{\rm 2B}(-2\mu)\epsilon_{\bf k}\right]^{1/2}$.
The most important feature of Eqs.~(\ref{h1}) and (\ref{h2}) is that
they contain no infrared and ultraviolet divergences and therefore
can be applied in any dimension and at all temperatures, even if
no condensate exists. 

Note that Eq.~(\ref{phase}) is also ultraviolet divergent. 
The ultraviolet divergences are removed by the same 
renormalization of the bare 
interaction $V_0$ and the final result is 
\bqa\nonumber
\langle\hat{\chi}({\bf x})\hat{\chi}({\bf x})\rangle
&=&{T^{\rm 2B}(-2\mu)\over V}\sum_{\bf k}\Bigg[
{1\over2\hbar\omega_{\bf k}}
\left[1+2N(\hbar\omega_{\bf k})\right]
\\&&
-{1\over2\epsilon_{\bf k}+2\mu}\Bigg]\;.
\label{h33}
\eqa
We will return to the physics of this important expression in 
Sec.~\ref{impph} below.

\subsection{Many-body T-matrix}
\label{mb}
In the previous section, we presented the modified Popov theory that takes
the phase fluctuations into account exactly. The final results in 
Eqs.~(\ref{h1}), (\ref{h2}), and (\ref{h33})
involve the two-body T-matrix. The two-body T-matrix takes into account
successive two-body scattering processes in vacuum.
However, it neglects the many-body effects of the surrounding gas.
In order to take this into account as well, we must use the many-body T-matrix 
instead of the two-body T-matrix in 
Eqs.~(\ref{h1}),~(\ref{h2}) and~(\ref{h33}).
Many-body effects have been shown to be 
appreciable in three dimensions only very close to 
the transition temperature~\cite{rgpaper}, 
but turn out to be much more 
important in one and two dimensions~\cite{henk1}.
Since the effects of the medium on the scattering properties of the atoms 
is only 
important at relatively high temperatures, we can apply a Hartree-Fock 
approximation to obtain for the many-body T-matrix
\bqa\nonumber
T^{\rm MB}(-2\mu)&=&T^{\rm 2B}(-2\mu)
\\&&
\hspace{-2cm}\times
\left[1+T^{\rm 2B}(-2\mu){1\over V}\sum_{\bf k}
{N(\epsilon_{\bf k}+n_0T^{\rm MB}(-2\mu))\over\epsilon_{\bf k}+\mu}
\right]^{-1}
\;.
\label{mbeq}
\eqa

The situation is in fact slightly more complicated because we
actually have two coupling constants in the equation for the chemical 
potential, which is the homogeneous version of the Gross-Pitaevskii equation.
When two atoms in the condensate collide at zero momentum, they both 
require an energy $\mu$ to be excited from the condensate, 
and thus the coupling is evaluated at $-2\mu$. 
This is the coupling that multiplies $n_0$ in the Gross-Pitaevskii equation.
On the other hand, the coupling that multiplies $n^{\prime}$
in the Gross-Pitaevskii equation involves one condensate atom and one atom 
in the thermal cloud, 
so that this coupling should now be evaluated at $-\mu$. 
The equation for the chemical potential thus becomes 
\begin{equation}
\mu=2n^\prime T^{\rm MB}(-\mu)+n_0T^{\rm MB}(-2\mu)
\label{mu100}.
\end{equation}
Note that the existence of two different many-body coupling constants 
for the interatomic interactions 
has previously been discussed by Proukakis {\it et al.} \cite{nick1}.
(See however also Ref.~\cite{snoopy}.)
This lead these authors to the so-called $G1$-theory, 
which is qualitatively somewhat similar to 
Eq.~(\ref{mu100}) but differs in detail.

\subsection{Long-wavelength physics}\label{impph}
We have given physical arguments for how to identify and subtract the
contribution to Eqs.~(\ref{1ln}) and~(\ref{1lmu}) from the phase fluctuations
of the condensate. At this point, we would like to give a somewhat more
rigorous field-theoretical argument. 
The Euclidean action that corresponds to the Hamiltonian in Eq.~(\ref{ham}) is
\bqa\nonumber
S[\psi^*,\psi]&=&
\int_0^{\hbar\beta}d\tau
\int d{\bf x}\;
\psi^*
\Bigg[
\hbar{\partial\over\partial\tau}
-{\hbar^2\over2m}\nabla^2
-\mu
\Bigg]  
\psi
\\
&-&{1\over2}\int_0^{\hbar\beta}d\tau \int d{\bf x}
V_0|\psi|^4
\label{ac}\;.
\eqa
If we substitute 
$\psi({\bf x},\tau)=\sqrt{n+\delta n({\bf x},\tau)}e^{i\chi({\bf x},\tau)}$ 
into
Eq.~(\ref{ac}), we obtain the action
\bqa\nonumber
S[\delta n,\chi]
&=&\int_0^{\hbar\beta}d\tau\int d{\bf x}
\Bigg[{1\over2}
\hbar{\partial \delta n \over\partial\tau}
\\&&\nonumber
+i\hbar(n+\delta n){\partial\chi\over\partial\tau}
+{\hbar^2\over2m}n\left(\nabla\chi\right)^2
\\&&
+{1\over2}\delta n\left({\hbar^2\over2m n}\nabla^2+V_0\right)
\delta n\Bigg]\;.
\label{x100}
\eqa
Here, $n$ is the average total density of the gas and
$\delta n({\bf x},\tau)$ represents the fluctuations.
At zero temperature, this action is exact
in the long-wavelength limit, if
$\left(\hbar^2k^2/2m n+V_0\right)$
is replaced by $\chi_{nn}^{-1}({\bf k})$, where $\chi_{nn}({\bf k})$ is 
the exact static density-density correlation function.

By using the classical equation
of motion to eliminate the phase $\chi({\bf x},\tau)$, we obtain 
the following action for the density fluctuations $\delta n({\bf x},\tau)$:
\bqa\nonumber
S[\delta n]&=&\int_0^{\hbar\beta}d\tau\int d{\bf x}
\Bigg[
-{m\over n}
{\partial\delta n\over\partial\tau}\nabla^{-2}
{\partial\delta n\over\partial\tau}
\\
&&
+{1\over2}\delta n\chi_{nn}^{-1}(-i\nabla)\delta n
\label{dact}
\Bigg]\;.
\eqa
The density fluctuations are therefore determined by 
\bqa\nonumber
\langle
\delta\hat{n}({\bf x})\delta\hat{n}({\bf x}^\prime)\rangle
&=&{1\over V}\sum_{\omega_n, {\bf k}}
{n\epsilon_{\bf k}\over\beta}
\left[{1\over(\hbar\omega_n)^2+(\hbar\omega_{\bf k})^2}\right]
\\
&&
\times
e^{i{\bf k}\cdot({\bf x}-{\bf x}^\prime)}
\;,
\eqa
where $\omega_n=2\pi n/\hbar\beta$ are the even Matsubara frequencies and
$\hbar\omega_{\bf k}=\sqrt{n\epsilon_{\bf k}/\chi_{nn}({\bf k})}$.
Summing over these Matsubara frequencies, we obtain
\bqa\nonumber
\langle
\delta\hat{n}({\bf x})\delta\hat{n}({\bf x}^\prime)\rangle
&=&{1\over V}\sum_{\bf k}{n\epsilon_{\bf k}\over\hbar\omega_{\bf k}}
\left[1+2N(\hbar\omega_{\bf k})\right]
\\
&\times&e^{i{\bf k}\cdot({\bf x}-{\bf x}^\prime)}
\;.
\eqa

Similarly, by using the classical equation of motion for 
$\delta n({\bf x},\tau)$,
we obtain from Eq.~(\ref{x100}) the following action for the phase 
fluctuations:
\bqa\nonumber
S[\chi]&=&\int_0^{\hbar\beta}d\tau\int d{\bf x}
\Bigg[
\hbar^2{\partial\chi\over\partial\tau}\chi_{nn}(-i\nabla)
{\partial\chi\over\partial\tau}
\\&&
+{\hbar^2n\over 2m}\left(\nabla\chi\right)^2
\Bigg]\;.
\label{pact}
\eqa
From this action, it is straightforward to calculate the 
propagator for the field $\chi({\bf x},\tau)$ 
and thereby the correlation function
$\langle\hat{\chi}({\bf x})\hat{\chi}({\bf x}^\prime)\rangle$. The result is
\bqa\nonumber
\langle\hat{\chi}({\bf x})\hat{\chi}({\bf x}^\prime)\rangle
&=&{1\over V}\sum_{\bf k}{1\over\chi_{nn}({\bf k})}{1\over2\hbar\omega_{\bf k}}
\left[1+2N(\hbar\omega_{\bf k})\right]
\\
&&
\times e^{i{\bf k}\cdot({\bf x}-{\bf x}^\prime)}\;.
\label{prop}
\eqa
Setting ${\bf x}^\prime={\bf x}$, 
we recover Eq.~(\ref{h33}) in the long-wavelength limit, 
if we use $\chi_{nn}({\bf k})\simeq1/T^{\rm MB}(-2\mu)$ for the static
density-density correlation function in that limit.
It is important to mention that Eq.~(\ref{prop}) is often used for the
short-wavelength part of the phase fluctuations as well~\cite{2d,1d}.
This is, however, incorrect because it contains ultraviolet divergences
due to the fact that the above procedure neglects interaction terms between
density and phase fluctuations that are only irrelevant at 
large wavelengths. The correct short wavelength behavior is given in
Eq.~(\ref{h33}).

\section{comparison with Popov Theory}
\label{usacomparison}
We proceed to compare predictions based on Eqs.~(\ref{h1}), (\ref{h2}), 
and (\ref{h33}) 
with exact results in one dimension and results based on the Popov theory
in two and three dimensions.
We consider only the homogeneous case here and discuss the 
inhomogeneous Bose gas in Sec.~\ref{trap}

\subsection{One dimension}
To understand the physical meaning of the
quantity $n_0$ in Eqs.~(\ref{h1}) and (\ref{h2}), i.e., whether it is 
the quasicondensate density or the true condensate density, we must determine
the off-diagonal long-range behavior of the one-particle density
matrix. Because this is a nonlocal property of the Bose gas, the
phase fluctuations contribute and we find in the large-$|{\bf x}|$ limit
\bqa\nonumber
\langle\hat{\psi}^{\dagger}({\bf x})\hat{\psi}({\bf 0})\rangle
&\simeq&
n_0\langle e^{-i\left(\hat{\chi}({\bf x})-\hat{\chi}({\bf 0})\right)}
\rangle \\
&=&
n_0 e^{-{1\over2}\langle\left[\hat{\chi}({\bf x})-\hat{\chi}({\bf 0})\right]^2\rangle}
\;.
\label{d11}
\eqa
Using Eq.~(\ref{h33}), we obtain  
for the exponent 
in Eq.~(\ref{d11}) 
\bqa\nonumber
\langle \left[\hat{\chi}({\bf x})-
             \hat{\chi}({\bf 0}) \right]^2 \rangle
      &      = &{T^{\rm MB}(-2\mu)\over V} \sum_{\bf k}
               \Bigg[{1\over\hbar\omega_{\bf k}}
\left[1+2N(\hbar\omega_{\bf k})\right]
               \\
      & -&
                     {1\over\epsilon_{\bf k}+\mu} \Bigg]
\times
        \left[1-\cos({\bf k}\!\cdot\!{\bf x})\right]\;.
\label{usa1}
\eqa
Writing the sum over wave vectors ${\bf k}$
as an integral, the phase fluctuations 
at zero temperature can be written as
\bqa
\langle\left[\hat{\chi}({\bf x})-\hat{\chi}({\bf 0})\right]^2\rangle&=&
{1\over2\pi n_0\xi}\int_0^{\infty}dk\;{1-\cos(kx)\over k\sqrt{k^2+1}}\;, 
\label{ff}
\eqa
where $\xi=\hbar/[4mn_0T^{\rm 2B}(-2\mu)]^{1/2}$ is the correlation length.
Note that we have used that $T^{\rm MB}(-2\mu)=T^{\rm 2B}(-2\mu)$ at 
zero temperature and that the chemical potential, as we show shortly, 
is to a good approximation equal to $n_0T^{\rm 2B}(-2\mu)$.
The integration can be performed analytically and the result is
\bqa\nonumber
\langle\left[\hat{\chi}({\bf x})-\hat{\chi}({\bf 0})\right]^2\rangle&=&
{1\over2\pi n_0\xi}
\Bigg[
{\pi x\over2\xi}\mbox{$_{1}$}F_2(1/2;1,3/2;x^2/4\xi^2)
\\
&&
\hspace{-2cm}
\label{hyper}
-{x^2\over2\xi^2}\mbox{$_{2}$}F_3(1,1;3/2,3/2;2x^2/4\xi^2)
\Bigg]\;,
\eqa
where $_{i}F_j(\alpha_1,\alpha_2,...\alpha_i;\beta_1,\beta_2,...\beta_j;x)$ 
are hypergeometric functions.
In the limit $|{\bf x}|\rightarrow\infty$, Eq.~(\ref{hyper}) reduces to
\bqa
\label{phfluc}
\langle\left[\hat{\chi}({\bf x})-\hat{\chi}({\bf 0})\right]^2\rangle
\simeq
{1\over2\pi n_0\xi}\log(x/\xi)\;.
\eqa
Using Eq.~(\ref{phfluc}) we find that the one-particle density matrix 
behaves for $|{\bf x}|\rightarrow\infty$ as
\bqa
\label{exact}
\langle\hat{\psi}^{\dagger}({\bf x})\hat{\psi}({\bf 0})\rangle\simeq
{n_0\over(x/\xi)^{1/4\pi n_0\xi}}\;.
\eqa

A few remarks are in order. 
First, the asymptotic behavior of the one-particle density
matrix
at zero temperature  
proves that the gas is not
Bose-Einstein condensed and that $n_0$ should be
identified with the quasicondensate density. 
Second, in the weakly-interacting 
limit $4\pi n\xi\gg1$ the depletion is small, 
so that, to first approximation, we can use $n_0\simeq n$ in 
the exponent $\eta=1/4\pi n_0\xi$. 
Indeed, from Eqs.~(\ref{h1}) and (\ref{h2}) we 
obtain the following expression for 
the fractional depletion of the quasicondensate 
\bqa
{n-n_0\over n}={1\over4\pi n\xi}\left({\sqrt{2}\over4}\pi-1\right)\;.
\eqa
We see that the expansion parameter is $1/4\pi n\xi$
and, therefore, that the depletion is very small.
Keeping this in mind, Eq.~(\ref{exact})
is in complete agreement with the exact result obtained by
Haldane~\cite{haldane}. Note that our theory cannot 
describe the strongly-interacting case $4\pi n \xi \ll 1$, where
the one-dimensional Bose gas behaves as a 
Tonks gas
\cite{olshanii,GW}. 

Finally, our results show that at a nonzero temperature 
the phase fluctuations increase as $\langle \left[\hat{\chi}({\bf x}) -
             \hat{\chi}({\bf 0}) \right]^2 \rangle\propto|{\bf x}|$ 
             for large distances, and
thus that the off-diagonal one-particle density matrix vanishes exponentially.
Hence, at nonzero
temperatures not even a quasicondensate exists and we have to use
the equation of state for the normal state
to describe the gas, i.e.,
\bqa
n={1\over V}\sum_{\bf k} N(\epsilon_{\bf k} + \hbar\Sigma - \mu)\;,
\eqa
where the Hartree-Fock self-energy satisfies
\bqa
\hbar\Sigma=2nT^{\rm MB}(-\hbar\Sigma\;)\;,
\eqa
and the many-body T-matrix obeys 
\bqa\nonumber
&&T^{\rm MB}(-\hbar\Sigma) =T^{\rm 2B}(-\hbar\Sigma) \nonumber\\
&\times&\left[1+ 
T^{\rm 2B}(-\hbar\Sigma) {1\over V}\sum_{\bf k}
{N(\epsilon_{\bf k}+\hbar\Sigma-\mu))
\over\epsilon_{\bf k}+\hbar\Sigma/2}  
\right]^{-1}\;,
\label{mbeq2}
\eqa
Note that the last three equations for the description of the normal
phase of the Bose gas are again valid for an arbitrary number of dimensions.

\subsection{Two dimensions}
In analogy with Eq.~(\ref{ff}), we obtain for the phase fluctuations
in two dimensions at zero temperature
\bqa
\langle\left[\hat{\chi}({\bf x})-\hat{\chi}({\bf 0})\right]^2\rangle&=&
{1\over2\pi^2 n_0\xi}\int_0^{\infty}dk\;{1-\cos(kx)\over\sqrt{k^2+1}}\;, 
\label{ff2d}
\eqa
Therefore we now find in the limit $|{\bf x}|\rightarrow \infty$ that  
\bqa
\langle{\hat{\psi}}^\dagger({\bf x})\hat{\psi}({0})\rangle=n_0
\eqa
and $n_0$ is clearly the condensate density of the gas. 
However, at nonzero temperatures the correlation function behaves as 
\bqa
\langle
{\hat{\psi}^\dagger({\bf x})}\hat{\psi}({0})\rangle
\simeq{n_0\over\left(x/\xi\right)^{1/n_0\Lambda^2}}
\;,
\eqa
where $\Lambda=\sqrt{2\pi\hbar^2/mk_{\rm B}T}$ 
is the thermal de Broglie wavelength 
and $n_0$ corresponds again to the quasicondensate density.
At zero temperature, the fractional depletion 
of the condensate in the 
Popov approximation was first calculated by Schick~\cite{schick}. 
He obtained 
\bqa
{n-n_0\over n}&=&
{1\over4\pi}T^{\rm 2B}(-2\mu)\;,
\eqa
where the chemical potential satisfies
$\mu=n T^{\rm 2B}(-2\mu)$.
The corresponding result based on Eqs.~(\ref{h1}) and~(\ref{h2}) is
\bqa
{n-n_0\over n}&=&{1\over4\pi}\left(1-\ln2\right)T^{\rm 2B}(-2\mu)\;,
\eqa
where $\mu$ now satisfies Eq.~(\ref{h2}). In two dimensions, the 
depletion predicted by the Popov theory is thus too large by
a factor of approximately 3.

In a number of applications, we need to calculate many-body correlators.
For instance, in order to calculate how a quasicondensate
modifies the two-body relaxation constants of a spin-polarized two-dimensional 
Bose gas, we need to know $K^{(2)}(T)\equiv\langle
\hat{\psi}^{\dagger}({\bf x})\hat{\psi}^{\dagger}({\bf x})
\hat{\psi}({\bf x})\hat{\psi}({\bf x})
\rangle/2n^2$. This correlator was considered in Ref.~\cite{henk2}
using the many-body T-matrix theory with an appropriate cutoff to remove 
the infrared divergences. 
An exact treatment of the phase fluctuations leads however directly
to an infrared finite result as we show now.
Using the parametrization in Eq.~(\ref{paraxy}) for the 
annihilation operators, we obtain first of all
\bqa\nonumber
\langle
\hat{\psi}^{\dagger}({\bf x})\hat{\psi}^{\dagger}({\bf x})
\hat{\psi}({\bf x})\hat{\psi}({\bf x})
\rangle
&=&n_0^2 
+n_0\Big[\langle\hat{\psi}^{\prime}({\bf x})
\hat{\psi}^{\prime}({\bf x})\rangle
\\&& \nonumber
\hspace{-1cm}
+\langle\hat{\psi}^{\prime\dagger}({\bf x})
\hat{\psi}^{\prime\dagger}({\bf x})\rangle
+4\langle\hat{\psi}^{\prime\dagger}({\bf x})
\hat{\psi}^{\prime}({\bf x})\rangle
\Big] \\ \nonumber
&&\hspace{-1cm}
+2\langle\hat{\psi}^{\prime\dagger}({\bf x})
\hat{\psi}^{\prime}({\bf x})\rangle^2\\&&
\hspace{-1cm}
+\langle\hat{\psi}^{\prime}({\bf x})\hat{\psi}^{\prime}({\bf x})\rangle
\langle\hat{\psi}^{\prime\dagger}({\bf x})
\hat{\psi}^{\prime\dagger}({\bf x})\rangle\;.
\label{corrdiv}
\eqa
The normal average is given by
$\langle \hat{\psi}'^{\dagger}({\bf x}) \hat{\psi}'({\bf x})
\rangle = n' + n_0 \langle \hat{\chi}({\bf x}) \hat{\chi}({\bf x})
\rangle$
and the anomalous average obeys $\langle \hat{\psi}'({\bf x})
\hat{\psi}'({\bf x}) \rangle = - n_0 \langle \hat{\chi}({\bf x})
\hat{\chi}({\bf x}) \rangle$, as we have seen. 
Using this, Eq.~(\ref{corrdiv}) can then be written as
\bqa\nonumber
\langle\hat{\psi}^{\dagger}({\bf x})\hat{\psi}^{\dagger}({\bf x})
\hat{\psi}({\bf x})\hat{\psi}({\bf x})\rangle
&=&n_0^2\left[1+2\langle\hat{\chi}({\bf x})\hat{\chi}({\bf x})
\rangle\right. \nonumber
\\&& \hspace{-2cm}\nonumber
\left.+3\langle\hat{\chi}({\bf x})\hat{\chi}({\bf x})\rangle^2\right] 
\\&&\hspace{-2cm}
+4n_0\left[1+
\langle\hat{\chi}({\bf x})\hat{\chi}({\bf x})\rangle\right]n^{\prime}
+2(n^{\prime})^2\;.
\eqa
Writing the correlator in this form, 
we explicitly see that the infrared divergences are due to spurious
contributions from the phase fluctuations.
Removing them, we obtain for the renormalized correlator 
\bqa
\label{corr1}
K^{(2)}_R(T)&=&{1\over2n^2}\left[n_0^2+4n_0n^{\prime}+
2\left(n^{\prime}\right)^2\right]\;.
\eqa

We would like to point out that critical fluctuations are not treated
within our mean-field theory. This is of course essential in the study
of the Kosterlitz-Thouless phase transition and we return to this issue 
in Sec.~\ref{kos}. However, an
example of a physical observable where
phase fluctuations are not important, is  
the three-body recombination rate constant.
We are at this point, therefore, already in the position 
to determine the reduction of the three-body
recombination rate constant due to the presence of a quasicondensate.
This can be expressed as~\cite{henk2}
\begin{equation}
{ L^{\rm N}\over L(T)} \simeq \left\{\left[{T^{\rm 2B}(-2\mu)\over
      T^{\rm 2B}(-2\hbar\Sigma)}\right]^6 K_R^{(3)}(T)\right\}^{-1}\;,
\end{equation}
where $L^{\rm N}$ is the recombination rate constant in the
normal phase, which is essentially independent of temperature, and
the self-energy satisfies 
$\hbar\Sigma=2nT^{\rm 2B}(-\hbar\Sigma)$. The renormalized three-body
correlator 
\begin{equation}
\label{corr}
K_R^{(3)}(T)={1\over6n^3}\Big[n_0^3 + 9n_0^2 n' + 18n_0(n')^2
                                + 6(n')^3\Big]
\end{equation}
is obtained from the expression for the correlation function
$\langle \hat{\psi}^{\dagger}({\bf x})
    \hat{\psi}^{\dagger}({\bf x}) \hat{\psi}^{\dagger}({\bf x})
    \hat{\psi}({\bf x}) \hat{\psi}({\bf x}) \hat{\psi}({\bf x}) \rangle$
by removing, as before, the spurious contributions from the phase
fluctuations. Moreover,
in two dimensions the T-matrix  depends logarithmically
on the chemical potential as
\begin{equation}
T^{\rm 2B}(-2\mu)= {4\pi\hbar^2\over m}
{1\over\ln(2\hbar^2/\mu m a^2)}\;,
\end{equation}
where $a$ is the two-dimensional $s$-wave scattering length.
In the case of atomic hydrogen adsorbed on a superfluid helium film, 
the scattering length was 
found to be $a=2.4a_0$~\cite{henk3}, where $a_0$ is the Bohr radius. 
However, there is some uncertainty
in this number because the hydrogen wave function
perpendicular to the helium surface is not known very 
accurately. In order to compare with experiment, we may therefore allow $a$
to vary somewhat.

\begin{figure}[htb]
\begin{center}
\epsfysize=9cm
\epsffile{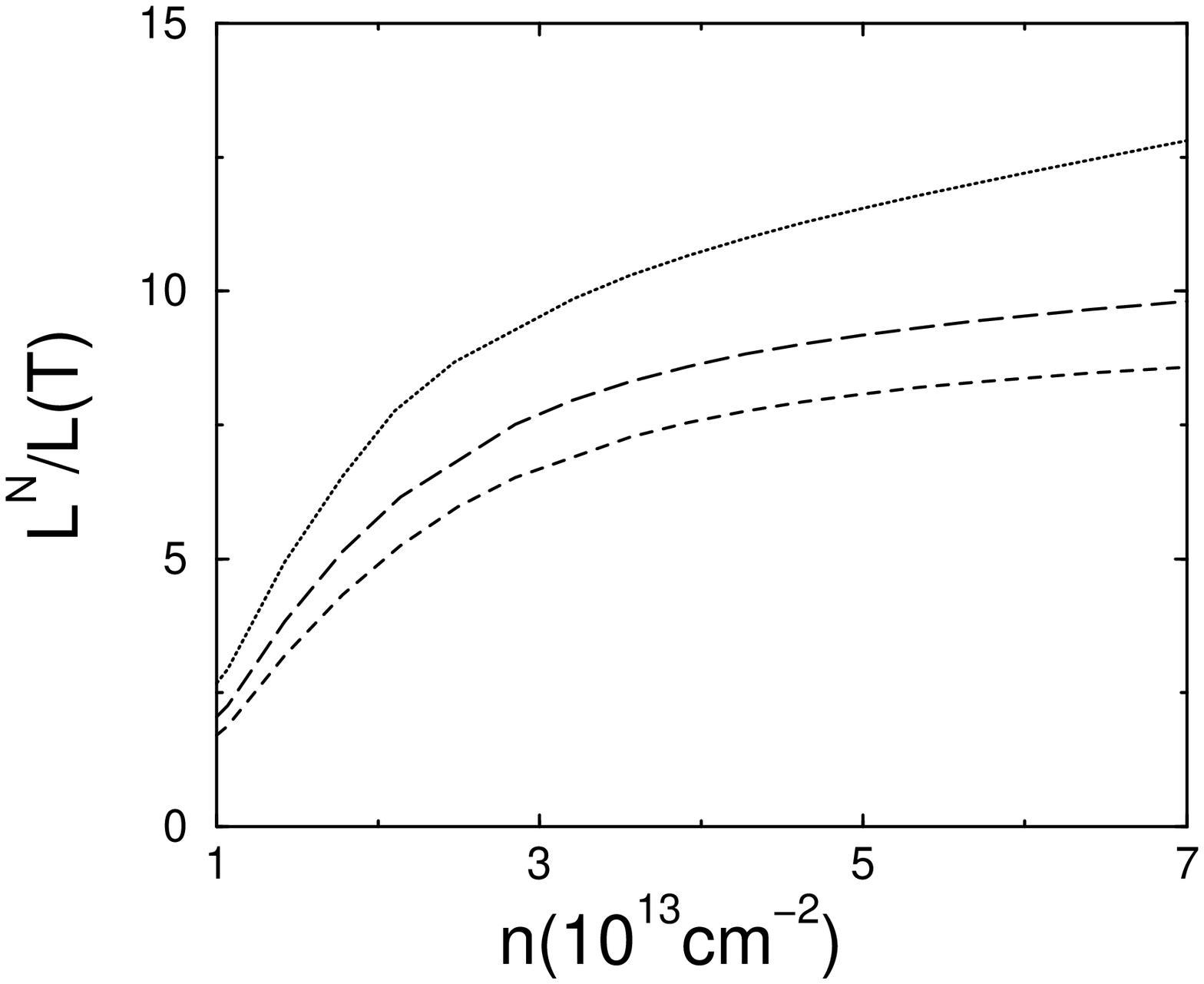}
\caption{
Reduction of the three-body recombination rate as a function of
the density for a temperature of $T=190$ mK and
three different values of the scattering length.
The dotted line corresponds to $a=2.4a_0$,
the long-dashed line to $a=1.2a_0$,
and the dashed line to $a=0.6a_0$, respectively.}
\label{fig1}
\end{center}
\end{figure}

In Fig.~\ref{fig1}, we show the reduction of the three-body
recombination rate as a function of the density at 
a fixed temperature $T=190$ mK for three different values of $a$. 
As can clearly be seen from Fig.~\ref{fig1}, the reduction of the 
three-body recombination rate is very sensitive to the value of $a$. 
What is most important at this point is that at high densities our 
calculation shows that the reduction of the recombination rate 
is much larger than the factor of 
6 predicted by Kagan {\it et al.}~\cite{kagan}. 
Such large reduction rates are indeed observed experimentally~\cite{saf}. 
A direct comparison, however, between the results of our theory and the 
measurements of Safonov {\it et al.} cannot be made here, since 
the density and temperature of 
the adsorbed hydrogen gas were not measured directly, but 
inferred from the properties of the three-dimensional buffer gas. 
Because this procedure requires knowledge of the 
equation of state of the two-dimensional Bose gas 
absorbed on the superfluid helium film, 
the raw experimental data needs to be reanalyzed 
with the theory presented in this paper. 
We can, however, compare the density at which the recombination 
rate 
starts to deviate considerably from the result in the normal state. 
For the temperature of $T=190$ mK, where most of the experimental data is 
taken, this is at a density of about $1.0\times10^{13}$ cm$^{-2}$, which is in 
excellent agreement with experiment. 
In view of this and the above mentioned problems we thus conclude 
that our results present a compelling theoretical explanation 
of the experimental findings.

\subsection{Three dimensions}
The Popov theory has been very successful in describing the 
properties of dilute three-dimensional trapped Bose gases.
It is therefore important to check that an exact treatment of the
phase fluctuations leads at most to small changes in the predictions for the 
three-dimensional case.

At zero temperature, the fractional depletion within the Popov theory
was first calculated by Lee and Yang~\cite{leeyang} and is given by
\bqa
{n-n_0\over n}&=& {8\over3}\sqrt{na^3\over\pi}\;,
\label{dep2}
\eqa%
where $a$ is the $s$-wave scattering length and we have used
\bqa
T^{\rm 2B}(-2\mu)&=&{4\pi a\hbar^2\over m}\;.
\eqa
The result that follows from Eqs.~(\ref{h1}) and~(\ref{h2}) is
\bqa
{n-n_0\over n}&=&
\left({32\over3}-2\sqrt{2}\pi\right)\sqrt{na^3\over\pi}\;.
\label{dep1}
\eqa
The fractional depletion is approximately 2/3 of the value
obtained from the Popov theory. 
It turns out that this is the largest change in the condensate depletion,
since the effects of phase fluctuations decrease with increasing temperature.
The critical temperature $T_{\rm BEC}$ is found by taking the limit
$n_0\rightarrow 0$ in Eqs.~(\ref{h1}) and~(\ref{h2}).
These expressions then reduce to the 
same expressions for the density and chemical
potential as in the Popov theory.
This implies that our critical temperature for Bose-Einstein
condensation coincides with  
that obtained in
Popov theory, i.e., the ideal gas result
\bqa
T_{\rm BEC}&=&{2\pi\hbar^2\over mk_{\rm B}}\left[{n\over\zeta\left(
\mbox{$3\over2$}
\right)}\right]^{2/3}\;,
\eqa
where $\zeta\left(
\mbox{$3\over2$}\right)\simeq2.612$.

\section{Kosterlitz-Thouless Phase Transition}
\label{kos}
In the previous section, we have compared our results using
the modified many-body T-matrix 
theory with established results in one, two, and three dimensions 
in the Popov approximation. 
Due to the mean-field nature
of the modified many-body T-matrix theory, the Kosterlitz-Thouless transition
is absent and a nontrivial solution of the equation of state
exists even if the superfluid density $n_s$ obeys $n_s\Lambda^2<4$. 
In this section, we correct for this by explicitly including the 
effects of vortex pairs in the phase fluctuations.
The idea is to use the modified many-body T-matrix 
theory to determine the initial values
of the superfluid density and the vortex fugacity, and to carry out
a renormalization-group calculation 
to find the fully renormalized values of these quantities. 
In this manner we can for example calculate
the critical temperature $T_c$ for the Kosterlitz-Thouless transition
given the scattering length $a$ and density $n$.

Let us for completeness first briefly sketch the derivation of the 
renormalization group equations for the superfluid density and the vortex 
fugacity \cite{girvinprivate}. 
Consider the velocity field of a vortex where the core is centered at the
positions ${\bf x}_i$, which we for simplicity take to lie on a 
lattice with an area of the unit cell equal to $\Omega$.
By rotating the velocity field by ninety degrees
we can map it onto the electric field of a point charge in two dimensions.
Since the total energy in both systems is proportional to the square
of the field integrated over space, there is complete analogy between 
a system of vortices and a two-dimensional Coulomb gas.
This analogy is very useful and we will take advantage of it in the following.
The total vorticity corresponds to the total charge of the Coulomb gas.
For the analogous two-dimensional neutral Coulomb gas on a square lattice,
the partition function can be written as
\bqa
Z=\sum_{\left\{{\bf x}_i,n_i\right\}}e^{-\beta\Big(\sum_{i\neq j}V({\bf x}_i
-{\bf x}_j)n_in_j
-E_c\sum_jn_j^2\Big)}\;,
\eqa
where $V({\bf x}_i-{\bf x}_j)=-2\pi\hbar^2n_s\ln(|{\bf x}_i-{\bf x}_j|/\xi$)/m 
is the Coulomb interaction between
two unit point charges in two dimensions, $n_s$ is the superfluid
density, and $E_c$
is the energy associated with the spontaneous creation of a charge,
i.e., it is the core energy of the vortices. 
The summation is over all possible configurations of charges $n_i$ at 
positions ${\bf x}_i$ on the lattice. 
The partition function can be rewritten in a field-theoretic fashion 
in terms of the electrostatic potential $\phi({\bf x})$ and the
fugacity $y=e^{-\beta E_c}$ as
\bqa\nonumber
Z&=&\sum_{\left\{{\bf x}_j,n_j\right\}}
\int\;{\cal D}\phi
e^{-\int d{\bf x}{1\over2}K^\prime(\nabla\phi({\bf x}))^2}
\\&& 
\times e^{-i\beta\Sigma_{j}n_j\phi({\bf x}_j)}
y^{\Sigma_jn_j^2}\;, 
\eqa
where $K^\prime=(2\pi)^2m/\hbar^2k_{\rm B}Tn_s$. 
In the limit where $y\ll 1$, the charge density 
is very low, and thus only $n_j=0, \pm1$ contribute to the partition function.
We can then write
\bqa\nonumber
Z&\simeq&
\int\;{\cal D}\phi e^{-\int d{\bf x}
{1\over2}K^\prime\left(\nabla\phi\right)^2}
\prod_j\Big[1+y\exp{(i\beta\phi({\bf x}_j))}
\\&&
\nonumber
              +y\exp{(-i\beta\phi({\bf x}_j))}+\dots\Big]
\\
&\simeq&
\int{\cal D}\phi\;e^{-\int d{\bf x}\left[
{1\over2}K^\prime\left(\nabla\phi\right)^2
-g\cos(\beta\phi)\right]}\;,
\label{ssg}
\eqa
where $g=2y/\Omega$. 
It is convenient to introduce a dimensionless dielectric 
constant $K$ that is related to $K^\prime$ by
$K=\beta^2/4\pi^2K^\prime=n_s\Lambda^2/2\pi$ where 
$\Lambda$ is the thermal wavelength.

The renormalization group equations for $K$, which is thus proportional to
the superfluid density,
and the fugacity $y$ 
can now be obtained by performing the usual momentum-shell integrations.
For the Sine-Gordon model derived in Eq.~(\ref{ssg}), this results in
\bqa
\label{rg1}
{dK^{-1}(l)\over dl}&=&4\pi^3y^2(l)+O(y^3)\;, \\
{dy(l)\over dl}&=&\left[2-\pi K(l)\right]y(l)+O(y^2)\;
\label{rg2}.
\eqa
The renormalization group equations to leading order in the variables 
$K(l)$ and $y(l)$ were first obtained by
Kosterlitz~\cite{koster2}, while the next-to-leading order terms were 
derived by
Amit {\it et al.}~\cite{amit}. The flow equations are not significantly
changed by including the higher-order corrections and we do not include
them in the following.

\begin{figure}[ht]
\epsfysize=9cm
\begin{center}
\centerline{\epsffile{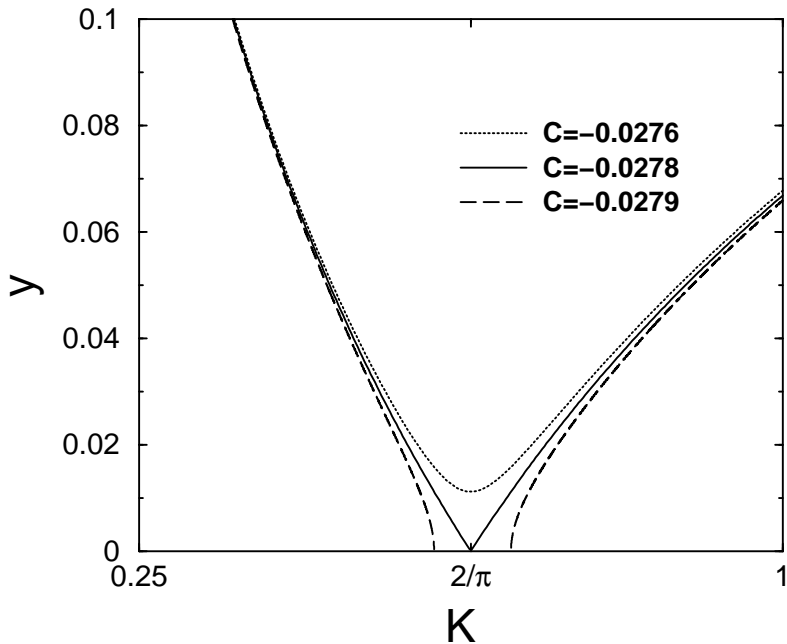}}
\end{center}
\caption[a]{
Renormalization group flow for the coupling constants $y$ and $K$. 
These curves are given by Eq.~(\ref{solu}) for different values of $C$.}
\label{fig2}
\end{figure}

The renormalization group equations~(\ref{rg1}) and~(\ref{rg2})
can be solved analytically by separation of variables and the solution is
\bqa
\label{solu}
y^2(l)-{1\over2\pi^3}\left[{2\over K(l)}+\pi\log(K(l))\right]=C\;,
\eqa
where the integration constant $C$ is determined by the initial conditions. 
For the critical trajectory it can be calculated by evaluating
the left-hand side at the fixed point $(y(\infty),K(\infty))=(0,2/\pi)$.
In this manner, we find $C=[\log(\pi/2)-1]/2\pi^2\simeq-0.0278$.
In Fig~\ref{fig2}, we show the flow of the Kosterlitz renormalization group 
equations.
There is a line of fixed points 
$y(\infty)=0$ and $K(\infty)\geq0$. The fixed point 
$(y(\infty),K(\infty))=(0,2/\pi)$ corresponds to the 
critical condition for the 
Kosterlitz-Thouless transition,
where the vortices start to unbind and superfluidity disappears. 
Physically this can be understood from the fact that below the transition the 
fugacity renormalizes 
to zero, which implies that at the largest length scales single vortices 
cannot be 
created by thermal fluctuations. They are therefore forced to occur in pairs.

The initial conditions for the renormalization group equations are
\bqa
K(0)&=&{\hbar^2n_0\over m k_{\rm B}T}\;,\\
y(0)&=& e^{-\beta E_c}\;,
\eqa
where $n_0$ is the quasicondensate
density and $E_c$ is the core energy of a vortex. Both are 
obtained from the modified many-body T-matrix theory considered previously. 
Writing the order parameter for a vortex configuration as $\psi_0({\bf x})=
\sqrt{n_0}f(x/\xi)e^{i\vartheta}$, where $\vartheta$ is the 
azimuthal angle, the core energy 
of a vortex follows from the Gross-Pitaevskii energy functional. It reads
\bqa
E_c&=&{\hbar^2\over2m}n_0\pi\int_0^{\infty}{dx}\;x\left[
\left(1-f^2\right)^2+2\left({df\over dx}\right)^2
\right]\;.
\eqa
The dimensionless integral was evaluated by Minnhagen and Nyl\'en, and 
takes the value 1.56~\cite{minn1}.

\begin{figure}[ht]
\epsfysize=9cm
\centerline{\epsffile{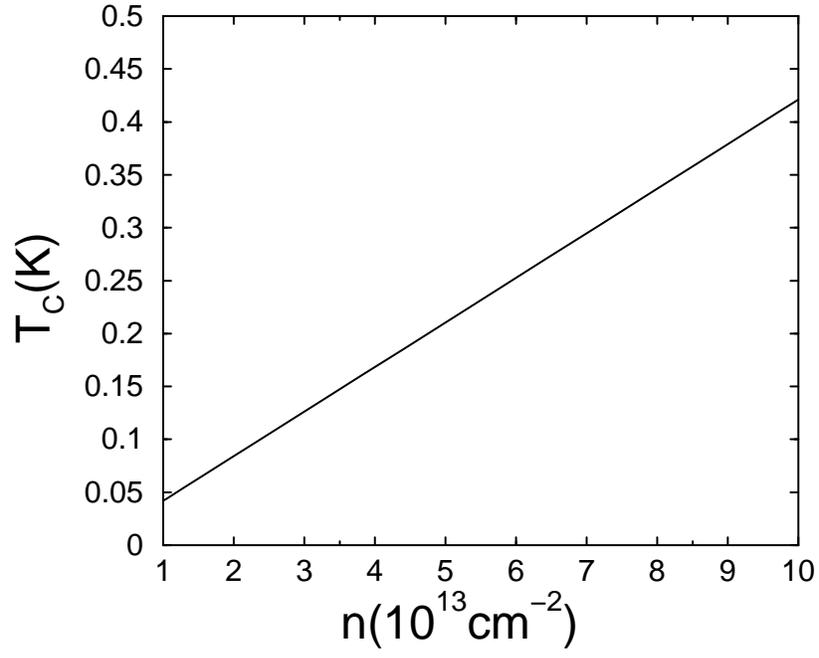}}
\caption[a]{
The critical temperature for the Kosterlitz-Thouless transition 
as a function of the density for 
spin-polarized atomic hydrogen with $a=2.4a_0$.}
\label{fig3}
\end{figure}

\begin{figure}[htb]
\begin{center}
\epsfysize=9cm
\centerline{\epsffile{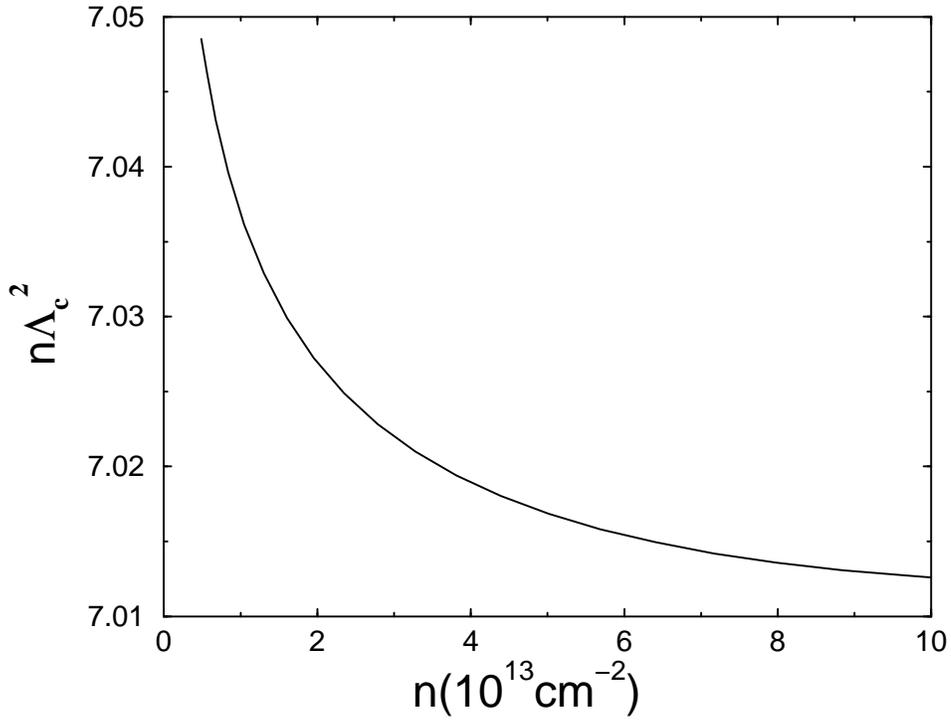}}
\end{center}
\caption[a]{
The critical degeneracy parameter $n\Lambda_c^2$ as a function of the density
for spin-polarized atomic hydrogen with $a=2.4a_0$. 
}
\label{fig4}
\end{figure}

Using the solution to the flow equations~(\ref{solu}) and the initial 
conditions, we can calculate the temperature for the Kosterlitz-Thouless
transition given the scattering length $a$ and the density of the system.
In the following, we consider again atomic hydrogen.
In Fig.~\ref{fig3}, we show the critical temperature as a function of density 
for $a=2.4a_0$.  We see that the critical temperature is essentially 
proportional to the density of the system. 
This can be seen in more detail in Fig.~\ref{fig4}, 
where we plot $n\Lambda_c^2$ as a function of $n$. It is 
clear from this figure that $n\Lambda_c^2$ indeed changes only 
slightly over the density range considered.

\begin{figure}[htb]
\begin{center}
\epsfysize=9cm
\centerline{\epsffile{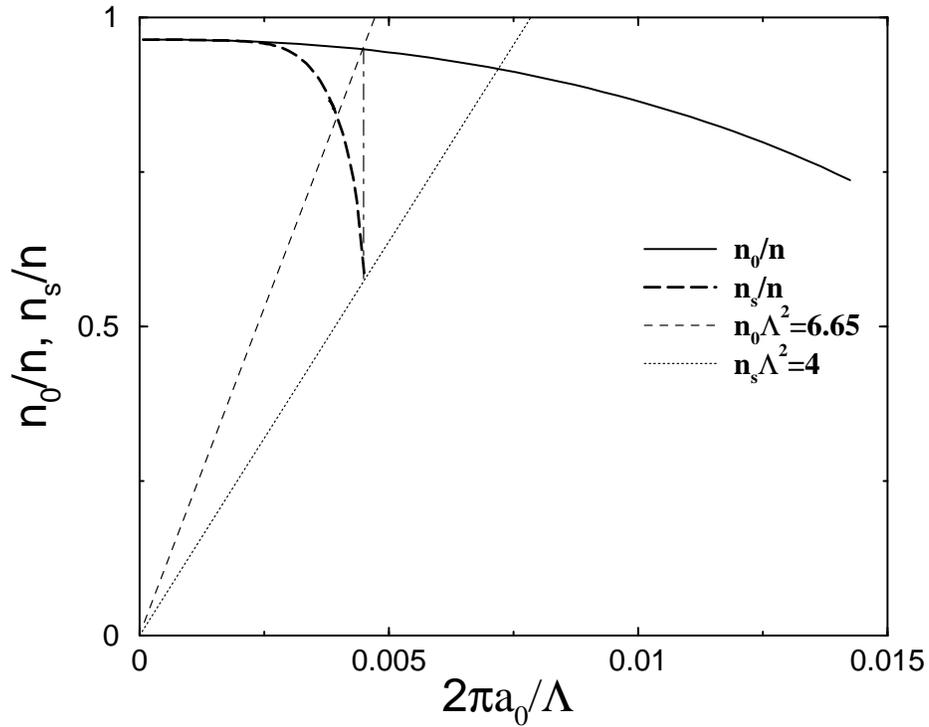}}
\end{center}
\caption[a]{
Quasicondensate density $n_0$ (solid curve) and 
superfluid density $n_s$ (long dashed curve) as a function of temperature. 
Also plotted are the Kosterlitz-Thouless condition $n_s\Lambda^2=4$ 
(dotted line) 
and the condition $n_0\Lambda^2=6.65$ (dashed line). 
The Kosterlitz-Thouless transition takes place when the dashed line 
intersects the 
solid curve. At the intersection point the long-dashed curve reaches the 
dotted line.
} 
\label{fig5}
\end{figure}

To understand the physics of the calculation 
better, we show in Fig.~\ref{fig5} the quasicondensate 
fraction $n_0/n$ following from the many-body $T$-matrix theory 
as a function of temperature for a total
density $n=1.25\times10^{13}$cm$^{-2}$. 
In addition, we show the superfluid density $n_s$ as calculated from the 
renormalization-group procedure explained previously.  
The Kosterlitz-Thouless transition takes place when $n_s$ lies on the line 
given by $n_s\Lambda^2=4$. 
Noticing that the left-hand side of Eq. (\ref{solu}) is a function of 
$n_0\Lambda^2$ only and solving the equation with respect to  
$n_0\Lambda^2$ using the value of $C$ at the transition, we 
obtain the condition $n_0\Lambda^2\simeq6.65$ 
for the Kosterlitz-Thouless transition.
It is therefore also seen in Fig.~\ref{fig5} that the Kosterlitz-Thouless 
transition takes place when 
the line given by $n_0\Lambda^2\simeq6.65$ intersects with the curve for $n_0$. 


\section{Trapped Bose Gases}\label{trap}
In this section, we generalize the theory presented in Secs. 
\ref{usapopov} and \ref{usacomparison} 
to inhomogeneous Bose gases. 
We also apply the results to a trapped one-dimensional Bose gas. 
We start by generalizing our previous expressions for the 
total density, Eq.~(\ref{h1}), and the phase fluctuations, 
Eq.~(\ref{h33}), to the inhomogeneous case.  
To do so we first consider the Gross-Pitaevskii equation
\bqa
&&\Bigg[
-{\hbar^2\over 2m}\nabla^2
+V^{\rm ext}({\bf x}) 
+T^{\rm MB}(-2\mu({\bf x}))|\psi_0({\bf x})|^2\nonumber\\
&+& 2T^{\rm MB}(-\mu({\bf x}))n^{\prime}({\bf x})
\Bigg]\psi_0({\bf x})=\mu\psi_0({\bf x})\;,
\label{gpu}
\eqa
which generalizes Eq.~(\ref{h1}) to trapped Bose condensates. 
Here the local chemical potential equals $\mu({\bf x})
=\mu-V^{\rm ext}({\bf x})$. 
The noncondensed density $n^\prime$({\bf x}) is to be determined by solving 
the Bogoliubov-de Gennes equations
\bqa\nonumber
\hbar\omega_j u_j({\bf x})&=&
\left[
-{\hbar^2\over 2m}\nabla^2
+V^{\rm HF}({\bf x})-\mu
\right]u_j({\bf x})\\
&+&T^{\rm MB}(-2\mu({\bf x}))n_0({\bf x})v_j({\bf x}) 
\label{u} \;, \\\nonumber
-\hbar\omega_j v_j({\bf x})&=&
\left[
-{\hbar^2\over 2m}\nabla^2
+V^{\rm HF}({\bf x})-\mu
\right]
v_j({\bf x})\\
&+&T^{\rm MB}(-2\mu({\bf x}))n_0({\bf x})u_j({\bf x})
\label{v}\;,
\eqa
where $n_0({\bf x})=|\psi_0({\bf x})|^2$ and the Hartree-Fock potential 
$V^{\rm HF}({\bf x})$ is given by
\bqa\nonumber
V^{\rm HF}({\bf x})&=&V^{\rm ext}({\bf x})
+2T^{\rm MB}(-\mu({\bf x}))n^\prime({\bf x})
\\
&&
+2T^{\rm MB}(-2\mu({\bf x}))n_0({\bf x})\;.
\eqa
The functions $u_j$ and $v_j$ are the usual Bogoliubov particle and hole 
amplitudes respectively,
which are chosen to be real here. In some cases, 
for instance when $\psi_0$ describes a vortex, we cannot choose
these amplitudes real and our equations are easily generalized to 
incorporate that fact.

In terms of the Bogoliubov amplitudes, the expression for the total density in 
Eq.~(\ref{h1}) reads
\bqa\nonumber
n({\bf x})&=&n_0({\bf x})+
\sum_j
\Bigg[\left(
u_j({\bf x})+v_j({\bf x})\right)^2N(\hbar\omega_j)
\\\nonumber
&+&v_j({\bf x})(v_j({\bf x})+u_j({\bf x}))\\
&+&{T^{\rm MB}(-2\mu({\bf x}))n_0({\bf x})\over 2\epsilon_j
+2\mu({\bf x})}(\phi_j({\bf x}))^2
\Bigg]\;.
\label{ntotal}
\eqa
Here, $\phi_j$ is the large-$j$ or high-energy limit of $u_j$ which can be 
obtained by neglecting the interaction terms in Eq.~(\ref{u}), namely
\begin{equation}
\epsilon_j \phi_j({\bf x})=
\left(
-{\hbar^2\over 2m}\nabla^2
+V^{\rm ext}({\bf x})-\mu
\right)\phi_j({\bf x})\;.
\label{uu}
\end{equation} 
In the large-$j$ limit, we also have
\begin{equation}
v_j({\bf x})=-{T^{\rm MB}(-2\mu({\bf x}))n_0({\bf x})\over2\epsilon_j}
\phi_j({\bf x})\;.
\label{vv}
\end{equation}   
It is clear that the expression of Eq.~(\ref{ntotal}) for the total density is
ultraviolet finite 
since the second and third term cancel each other in the large-$j$ limit.

Finally, the phase fluctuations in the trapped case are determined by 
$\langle\hat{\chi}({\bf x})\hat{\chi}({\bf x}^{\prime})\rangle$ which is 
given by
\bqa
\nonumber
\langle\hat{\chi}({\bf x})\hat{\chi}({\bf x}^{\prime})\rangle
&=&\\ \nonumber
&&\hspace{-2cm}
-\sum_j
{1\over2\sqrt{n_0({\bf x})n_0({\bf x}^{\prime})}}
\Bigg\{
u_j({\bf x}^\prime)v_j({\bf x})\left[1+2N(\hbar\omega_j)\right]
\\&& \nonumber
\hspace{-2cm}
+\Bigg[{T^{\rm MB}(-2\mu({\bf x}))n_0({\bf x})
\over2\epsilon_j+2\mu({\bf x})}
\Bigg]
\phi_j({\bf x}^\prime)\phi_j({\bf x})
\\&& \nonumber
+u_j({\bf x})v_j({\bf x}^\prime)\left[1+2N(\hbar\omega_j)\right]
\\&& 
\hspace{-2cm}
+\Bigg[{T^{\rm MB}(-2\mu({\bf x}^{\prime}))n_0({\bf x}^{\prime})
\over2\epsilon_j+2\mu({\bf x}^{\prime})}
\Bigg]
\phi_j({\bf x})\phi_j({\bf x}^{\prime})
\Bigg\}\;.
\label{chi}
\end{eqnarray}
In particular the normalized form 
of the off-diagonal one-particle density matrix of 
Eq.~(\ref{d11}) becomes for large distances 
$|{\bf x}-{\bf x}^{\prime}|$ equal to
\bqa
g^{(1)}({\bf x}, {\bf x}^{\prime})&=&\exp\left(
-\langle[\hat{\chi}({\bf x})-\hat{\chi}({\bf x}^{\prime})]^2\rangle/2
\right)\;.
\label{corr11}
\eqa

\subsection{Density profiles}
\label{prof}
We are now ready to calculate the 
total density profile by solving
Eqs.~(\ref{gpu}) and~(\ref{ntotal}) selfconsistently. 
In the remainder of the paper, we restrict ourselves to one-dimensional 
harmonic traps 
with, therefore,
\begin{equation}
V^{\rm ext}(z)={1\over2}m\omega_z^2z^2\;.
\label{vext}
\end{equation}
For simplicity we use 
the local-density approximation, which allows us to 
calculate the densities directly
using the many-body generalization of Eq.~(\ref{h1}), and Eq.~(\ref{mu100}). 
In Fig.~\ref{fig6} the total density profile is shown at four
different values of the 
temperature. 
For the four different temperatures each of the four curves 
is composed of two parts. The first part near the center of the trap 
represents the superfluid part of the gas and contains the (quasi)condensate.  
The other part consists only of the noncondensed atoms. 
The small discontinuity between the two parts is caused by the use of 
two different equations of state for the superfluid and thermal phases of the gas. 
In the following we call the position 
of the discontinuity the temperature-dependent 
Thomas-Fermi radius of the (quasi)condensate.  
For distances below the discontinuity we use the above-mentioned equations,
while for
distances above the discontinuity, we simply use
\bqa
\nonumber
n(z)&=&\;\int_{-\infty}^{\infty}{dk\over2\pi}\; 
N\left(\epsilon_{k}+m\omega_z^2z^2/2\right.\\
&+&\left. 2nT^{\rm MB}(-\hbar\Sigma(z))-\mu\right)\;.
\eqa
For all these curves $\mu=30\hbar\omega_z$. The remaining parameters
used here are those of the experiment of 
G\"orlitz {\it et al.}~\cite{lowdketterle}. 
In particular, we have used $^{23}$Na in the trap with
$\omega_z=2\pi\times3.5$ rad/sec, 
$l_z=\sqrt{\hbar/m\omega_z}\simeq1.12\times10^{-5}$m.
The three-dimensional $s$-wave 
scattering length is $a\simeq2.75$ nm which is related to 
the 
one-dimensional scattering length $\kappa^{-1}$ 
defined by $T^{\rm 2B}(-2\mu)=4\pi\kappa\hbar^2/m$.
For harmonic confinement, we have 
$\kappa=a/2\pi l_\perp^2$, where $l_\perp$ is the harmonic oscillator 
length of the axially symmetric trap in the 
direction perpendicular to the $z$-axis. 
We have used $\omega_{\perp}=2\pi\times360$ rad/sec and 
$l_{\perp}=\sqrt{\hbar/m\omega_{\perp}}\simeq1.10\times10^{-6}$m.

As expected, the temperature-dependent Thomas-Fermi radius decreases with 
increasing the temperature. 
At the temperature when this radius vanishes the one-dimensional system 
reaches the crossover temperature for the formation of a (quasi)condensate. 
We have calculated this crossover temperature for different values of 
the scattering length at a  
constant value of the number of atoms, the latter being fixed
by adjusting the 
chemical potential. 
In Fig.~\ref{fig7}, we show the result of this calculation, and plot 
the crossover temperature $T_{\rm QC}$ and 
the chemical potential 
against the scattering length.
The inset in Fig.~\ref{fig7} shows that on a double logarithmic scale 
the temperature $T_{\rm QC}$  
is clearly not a straight 
line indicating that the relation between $T_{\rm QC}$ 
and $\kappa$ is not a simple power law and may contain logarithmic 
dependence. 
It is shown in Ref.~\cite{van} that for $a=0$, 
the transition temperature $T_{\rm QC}$ should satisfy 
$T_{QC}=N\hbar\omega_z/k_{\rm B}\ln{(2N)}$, 
where $N$ is the number of atoms. 
In the case of Fig.~\ref{fig7} we have $N=950$, which leads to 
$T_{QC}\simeq164T_0$ for an ideal gas. 
Of course, this limit is not obtained in Fg.~\ref{fig7} because our 
calculation is based on a 
local-density approximation, which will always break down for sufficiently 
small values of $\kappa$.
On the other hand, the curve for the chemical potential 
becomes almost a straight line on a double logarithmic scale. A calculation of 
the slope 
of this line shows that the slope starts at a value slightly larger than 2/3 
at the 
lower end of the curve and saturates at this value near the 
upper end. 
The value 2/3 is what we expect, since in the Thomas-Fermi limit 
it is easy to show that $\mu=(3\pi/\sqrt{2})^{2/3}(N\kappa )^{2/3}
\hbar\omega_z\simeq3.5(N\kappa )^{2/3}\hbar\omega_z$. 
Calculating similar curves for different values of $N$ 
we actually find numerically that $\mu\simeq3.2(N\kappa )^{2/3}\hbar\omega_z$.

\subsection{Phase fluctuations}

The aim of this section is to calculate the 
normalized off-diagonal density matrix given 
by Eq.~(\ref{corr11}). This function expresses the coherence in the system. 
It is calculated by solving the Bogoliubov-de Gennes equations in 
Eqs.~(\ref{u}) and 
(\ref{v}) using the density profile calculated in the previous subsection. 
Specifically 
from the (quasi)condensate density profile $n_0$, we determine a 
temperature-dependent 
Thomas-Fermi radius. This radius is then used to 
calculate 
the phase fluctuations at that specified temperature in the following manner. 

We start by employing the following scaling: lengths are scaled to the trap 
length $l_z=(\hbar/m\omega_z)^{1/2}$, frequencies to $\omega_z$, 
energies to $\hbar\omega_z$, 
and densities to 
$4\pi/l_z$. 
With this scaling, the Bogoliubov-de Gennes equations  
take the dimensionless form 
\begin{equation}
\omega_j u_j=
\left(
-{1\over 2}{d^2\over dz^2}+{1\over 2}z^2-\mu+2 \kappa n
\right)u_j-\kappa n_0 v_j\;,
\label{uscaled}
\end{equation}    
\begin{equation}
-\omega_j v_j=
\left(
-{1\over 2}{d^2\over dz^2}+{1\over 2}z^2-\mu+2 \kappa n
\right)v_j-\kappa n_0 u_j\;.
\label{vscaled}
\end{equation} 
Using the same scaling, the Gross-Pitaevskii equation takes the form 
\begin{equation}
\left[
-{1\over 2}{d^2\over dz^2}+{1\over 2}z^2-\mu
+\kappa(n_0+2n^{\prime})
\right]
\sqrt{n_0}=0\;.
\label{gpuscaled}
\end{equation}

Next we define $F_j(z)=u_j(z)+v_j(z)$ and $G_j(z)=u_j(z)-v_j(z)$, and  
derive from Eqs. (\ref{uscaled}) 
and (\ref{vscaled}) two equations for $F_j(z)$ and $G_j(z)$, namely
\bqa\nonumber
{d^4F\over dz^4}
-2(f+g){d^2F\over dz^2}
-4{dg\over dz}
{dF\over dz}
\\
-\Big(4\omega_j^2+2{d^2g\over dz^2}-4gf\Big)F&=&0\;,
\label{F} \\ \nonumber
{d^4G\over dz^4}
-2(f+g){d^2G\over dz^2}
-4{df\over dz}
{dG\over ds}
\\
-\Big(4\omega_j^2+2{d^2f\over dz^2}-4gf\Big)G&=&0\;,
\label{G}
\eqa
where 
the functions $f(z)$ and $g(z)$ are given by
\bqa
f&=&{1\over2}z^2+2\kappa n -\mu+\kappa n_0 \;,
\label{sf} \\
g&=&{1\over2}z^2+2\kappa n -\mu-\kappa n_0 \;.
\label{sg}
\eqa
For our purposes we can use the Thomas-Fermi approximation which 
neglects the derivative term in Eq.~(\ref{gpuscaled}). 
Hence  
\beq
\left[
{1\over 2}z^2-\mu
+\kappa(n_0+2n^{\prime})
\right]
\sqrt{n_0}=0\;.
\label{gpuscaledtf}
\eeq
In this limit, the functions $f(z)$ and $g(z)$ are given by 
$f(z)=2\kappa n_0(z)$ and $g(z)=0$. 
In the Thomas-Fermi approximation, 
we substitute 
these values for $f(z)$ and $g(z)$ 
into Eqs.~(\ref{F}) and~(\ref{G}) 
and neglect the fourth-order derivative terms. 
These Equations thus take the form
\bqa
\kappa n_0{dF_j^2\over dz^2}
+\omega_j^2F_j&=&0\;,
\label{Ftf}\\
{d^2(\kappa n_0G_j)\over dz^2}
+\omega_j^2G_j&=&0\;.
\label{Gtf}
\eqa
In Ref.~\cite{fetterusa}, it was shown 
that $\sqrt{\kappa n_0(z)}G_j(z)$ corresponds to 
density 
fluctuations and $F_j(z)/\sqrt{\kappa n_0(z)}$ 
corresponds to phase fluctuations in the
hydrodynamic approach~\cite{stringari}. 
We therefore define the function $h_{j}(z)$ 
\begin{equation}
h_j=\sqrt{\kappa n_0}G_j=F_j/\sqrt{\kappa n_0}\;.
\label{h}
\end{equation}
Substituting this 
back in Eqs.~(\ref{Ftf}) and~(\ref{Gtf}) both equations reduce 
to a single equation for $h_j(z)$, namely
\begin{equation}
\kappa n_0{d^2h_j\over dz^2}+\kappa {dn_0\over dz}{dh_j\over dz}
+\omega_j^2h_j=0\;.
\label{heq}
\end{equation}
This equation can finally be simplified 
using the Thomas-Fermi expression 
for $\kappa n_0(z)$ from Eq.~(\ref{gpuscaledtf}),
namely $\kappa n_0(z)\simeq\mu^{\prime}-z^2/2$ where 
$\mu^{\prime}=\mu-2\kappa n^\prime(0)$. 
Note that we have made the approximation that we take $n^\prime(z)$ to be equal 
to its value at the center, namely $n^\prime(0)$. 
This approximation is justified in view of the fact that the presence of the 
condensate repels the atoms from the noncondensate atoms from the center of 
the trap. This is also supported by a numerical solution of 
Eqs.~(\ref{h1}) 
and~(\ref{mu100}), where we find 
that $n^{\prime}(z)\ll n_0(z)$,
except at the Thomas-Fermi radius where they become of the same order. 
Moreover, the slope of $n^\prime(z)$ is small 
for distances close to 
the center. 
Thus, the last equation becomes
\begin{equation}
(1-y^2){d^2\over dy^2}h_j(y)-2y{d\over y}h_{j}(y)+2\omega_j^2h_j(y)=0\;,
\label{heqscaled}
\end{equation} 
where $y=z/R_{\rm TF}(T)$ and
$R_{\rm TF}(T)=\sqrt{2\mu^{\prime}(T)}$ is the Thomas-Fermi radius

In the following, we reinstate the units.
Interestingly, Eq.~(\ref{heqscaled}) is the Legendre equation with the 
Legendre polynomials 
as solutions:
\bqa
h_j(z)=P_j(z/R_{\rm TF})=P_j(y)\;,
\eqa
where the energy eigenvalues are
\begin{equation}
\hbar\omega_j=\sqrt{{j(j+1)\over2}}
\hbar\omega_z
\;,\hspace{2cm}j=0,1,2,\dots\;.
\label{dispersion}
\end{equation}
The normalization condition for the Bogoliubov amplitudes is
\begin{equation}
\int_{-R_{TF\rm }}^{R_{TF\rm }}
dz\;\left[|u_j|^2(z)-|v_j|^2(z)\right]=1\;,
\label{norm}
\end{equation}
which leads to
\bqa
F_j(z)&=&{1\over\sqrt{R_{\rm TF}}}\sqrt{(j+1/2)\mu^\prime\over\hbar\omega_j}
\sqrt{1-y^2}P_j(y)\;,
\label{Ffinal} \\
G_j(z)&=&{1\over\sqrt{R_{\rm TF}}}\sqrt{(j+1/2)\hbar\omega_j\over\mu^\prime}
{P_j(y)\over\sqrt{1-y^2}}\;.
\label{Gfinal}
\eqa
These expressions are in agreement with those obtained in
Ref.~\cite{ghora}. 
Consequently, we find
\bqa
u_j(z)&=&{1\over2}
\left(
A_j\sqrt{1-y^2}+{B_j\over\sqrt{1-y^2}}
\right)
P_j(y)\;,
\label{ufinal} \\
v_j(z)&=&{1\over2}
\left(
A_j\sqrt{1-y^2}-{B_j\over\sqrt{1-y^2}}
\right)
P_j(y)\;,
\label{vfinal}
\eqa
where 
\bqa
A_{j}&=&{1\over\sqrt{R_{\rm TF}}}\sqrt{{(j+1/2)\mu^\prime\over\hbar\omega_j}}\\
B_{j}&=&{1\over\sqrt{R_{\rm TF}}}\sqrt{{(j+1/2)\hbar\omega_{j}\over\mu^\prime}}\;.
\eqa 
The expression for the phase fluctuations in Eq.~(\ref{chi}) now reads, 
after neglect of the quantum contribution, 
\begin{eqnarray}\nonumber
\langle
\left[
\hat{\chi}(z)-\hat{\chi}(z^{\prime})
\right]^2
\rangle
&=& \\ \nonumber
&&
\hspace{-1cm}
{4\pi \kappa l_z^4\over R_{\rm TF}^2}
\sum_{j=0}N(\hbar\omega_j)
\Bigg\{
A_j^2\Big[P_j(y)-P_j(y^{\prime})\Big]^2
\\&&
\hspace{-1cm}
-B_j^2\left[{P_j(y)\over1-y^2}-{P_j(y^{\prime})\over1-y^{\prime 2}}\right]^2
\Bigg\}\;.
\label{chifinal}
\end{eqnarray}  
It should be noted 
that the first term in this sum, $j=0$,  does not diverge 
as one might think at first instance. It actually vanishes and
the sum can start from $j=1$.
Physically, this is a result of the fact that the global phase does not
influence the phase fluctuations. 

For the four values of temperature used in Fig.~\ref{fig6},
we insert the 
corresponding $R_{\rm TF}(T)$ in Eq.~(\ref{chifinal}) to calculate the 
phase correlation function $g^{(1)}(0,z)$.
In Fig.~\ref{fig8}, we plot this quantity and we see that at 
sufficiently 
low temperatures the phase correlation function decreases only 
slightly over the condensate 
size. This indicates that a true condensate can exist at sufficiently low 
temperatures for interacting trapped one-dimensional Bose gases.

\subsection{Comparison with exact results}
We next compare the above results to predictions
based on a Langevin field equation for the order parameter of a trapped, 
one-dimensional condensate 
in contact with a three-dimensional Bose gase that acts as a ``heat bath''. 
Such a situation can be created experimentally in a 
magnetically trapped three-dimensional system, by using a laser beam 
to provide and additional optical potential along two of the directions. 
The laser beam then needs to be focused such that the motion
of the system freezes out along these directions. 
The gas in the potential ``dimple'' provided by the laser then indeed becomes
an effectively one-dimensional condensate, in contact with the 
three-dimensional thermal cloud in the magnetic trap, which acts as its 
heat bath. The dynamics of the order 
parameter is governed in this case by~\cite{Stoof_Booklet,Stoof_Noisy}
\begin{eqnarray}\nonumber
i \hbar \frac{ \partial \Phi(z,t) }{ \partial t} 
&= & \Bigg[ - \frac {\hbar^{2} \nabla^{2} }{2m} + V^{\rm ext}(z) - \mu 
- iR(z,t) 
\\& & 
+ g |\Phi(z,t)|^{2} \Bigg] \Phi(z,t) + \eta(z,t)\;,
\label{lang}
\end{eqnarray}
where the external trapping potential in the weakly-confined
direction  $V^{\rm ext}(z)$ is again given in Eq.~(\ref{vext}) 
and $\mu$ is the effective chemical potential of the one-dimensional system. 
The one-dimensional coupling constant $g$ is related to $\kappa$ by 
$g=4\pi\kappa\hbar^2/m$.
Physically, the function $iR(z,t)$ describes the
pumping of the one-dimensional condensate from the surrounding thermal 
cloud, and $\eta(z,t)$ corresponds to the associated noise with Gaussian 
correlations.
Both these quantities depend on the one-dimensional Keldysh
self-energy $\hbar\Sigma^{K}(z)$, as discussed in detail in 
Ref.~\cite{Stoof_Noisy}.
For our purposes, we only need that
\bqa%
iR(z,t)&=&-{\beta\over4}\hbar\Sigma^{\rm K}(z)\nonumber
\\&&
\hspace{-2cm}
\times
\left(
-{\hbar^2\nabla^2\over2m}+V^{\rm ext}(z)-\mu+T^{\rm 2B}|\Phi(z,t)|^2
\right)\;, \\
\langle
\eta^*(z,t)\eta(z^\prime,t^{\prime})
\rangle
&=&{i\hbar^2\over2}\Sigma^{\rm K}(z)\delta(z-z^\prime)\delta(t-t^{\prime})\;,
\eqa
where $\langle...\rangle$ denotes averaging over the realizations 
of the noise $\eta(z,t)$.
The numerical techniques employed are discussed in 
Ref.~\cite{Stoof_Noisy}, where it also was shown that 
with the last two expressions,
the trapped gas relaxes to the correct equilibrium, as ensured by the 
fluctuation-dissipation theorem. To simplify the numerics, 
the noncondensed part in the dimple is here
allowed to relax to the ``classical'' value 
$N(\epsilon)=[\beta ( \epsilon- \mu )]^{-1}$, and the comparison 
to the previous mean-field predictions is therefore 
carried out by making the 
same approximation in the calculation of both $n_{0}(z)$ and $n^{\prime}(z)$.
The normalized first-order correlation function at equal time 
$g^{(1)}(0,z)$ corresponding to the previously computed phase
correlation function, is calculated via numerical
autocorrelation measurements, i.e., 
\begin{equation}
g^{(1)}(0,z,t) = \frac{ \langle \Phi^{*}(0,t) \Phi(z,t) \rangle }
{ \sqrt{ \langle |\Phi(0,t)|^{2} \rangle \langle |\Phi(z,t)|^{2} 
\rangle }}\;,
\end{equation}
where the brackets again denote averaging over the different 
realizations of the noise. Of course, the time $t$ must be sufficiently large
so that the gas has relaxed to thermal equilibrium and 
$g^{(1)}(0,z,t)$ is independent of time.

\begin{figure}[htb]
\epsfysize=9cm
\begin{center}
\centerline{\epsffile{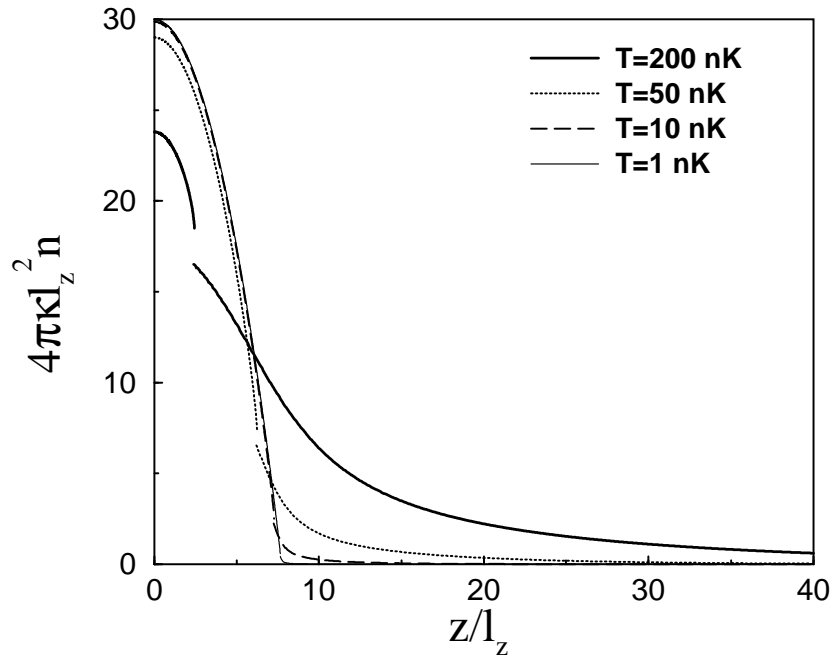}}
\end{center}
\caption[a]{
Density profile of a trapped one-dimensional Bose gas at four different
temperatures. The quantities $l_z$ and $\kappa$ are defined in the text. 
}
\label{fig6}
\end{figure}
\begin{figure}[ht]
\epsfysize=9cm
\begin{center}
\centerline{\epsffile{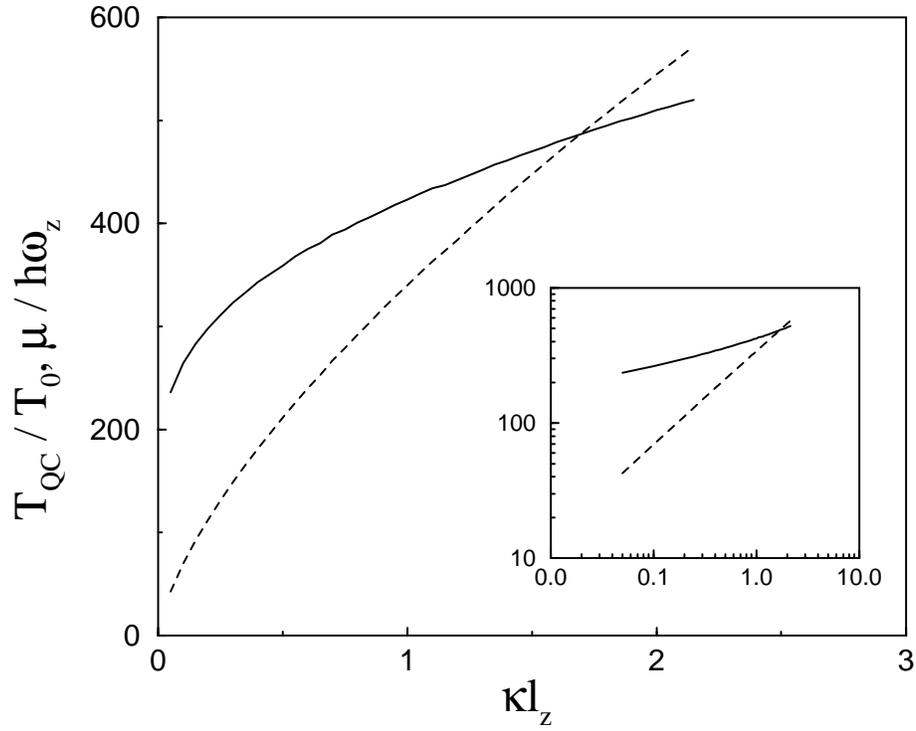}}
\end{center}
\caption[a]{
The crossover temperature 
$T_{\rm QC}$ is shown with the solid curve, 
and the chemical potential at this temperature is  
shown with the dashed curve, both as a function of the coupling constant. 
The temperature is scaled to $T_0=\hbar^2/mk_{\rm B}l_z^2$. 
The inset shows the same curves on a double logarithmic scale. 
}
\label{fig7}
\end{figure}
\begin{figure}[ht]
\epsfysize=9cm
\begin{center}
\centerline{\epsffile{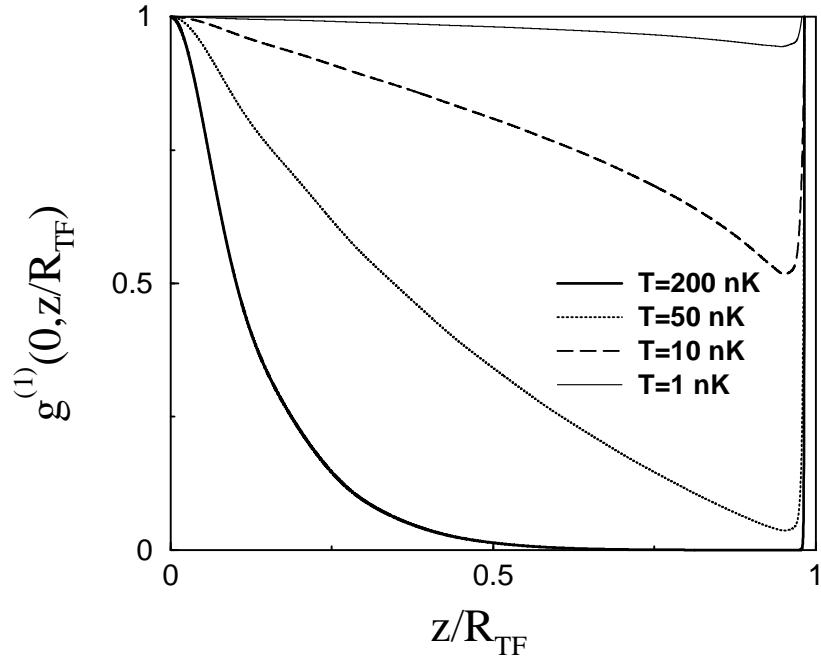}}
\end{center}
\caption[a]{
Normalized first-order (phase) correlation function as a function of 
position for different temperatures. 
}
\label{fig8}
\end{figure}
\begin{figure}[ht]
\epsfysize=9cm
\begin{center}
\centerline{\epsffile{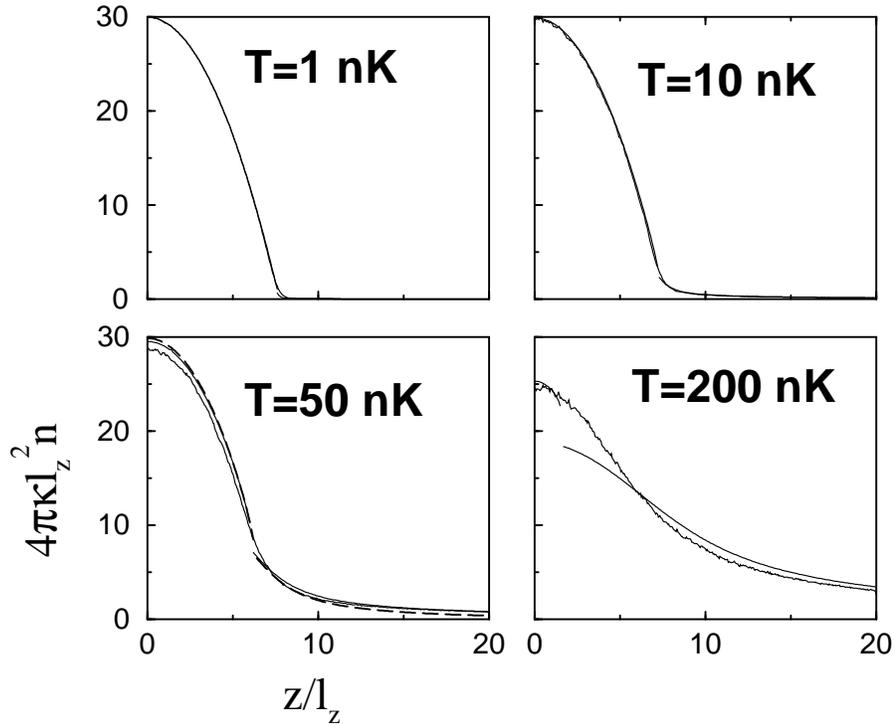}}
\end{center}
\caption[a]{
Comparison of the mean-field densities profiles   
(solid curves) to numerical solutions of the Langevin equation 
in Eq.~(\ref{lang}) (noisy curves). All the above 
curves are calculated using the classical approximation of 
the Bose-Einstein distribution function. 
For the $T=50$ nK case we have also 
plotted the corresponding density calculated using the full 
Bose-Einstein distribution function (dashed curve) in order to show
the difference between the classical and quantum mean-field approximations.} 
\label{fig9}
\end{figure}
\begin{figure}[htb]
\epsfysize=9cm
\begin{center}
\centerline{\epsffile{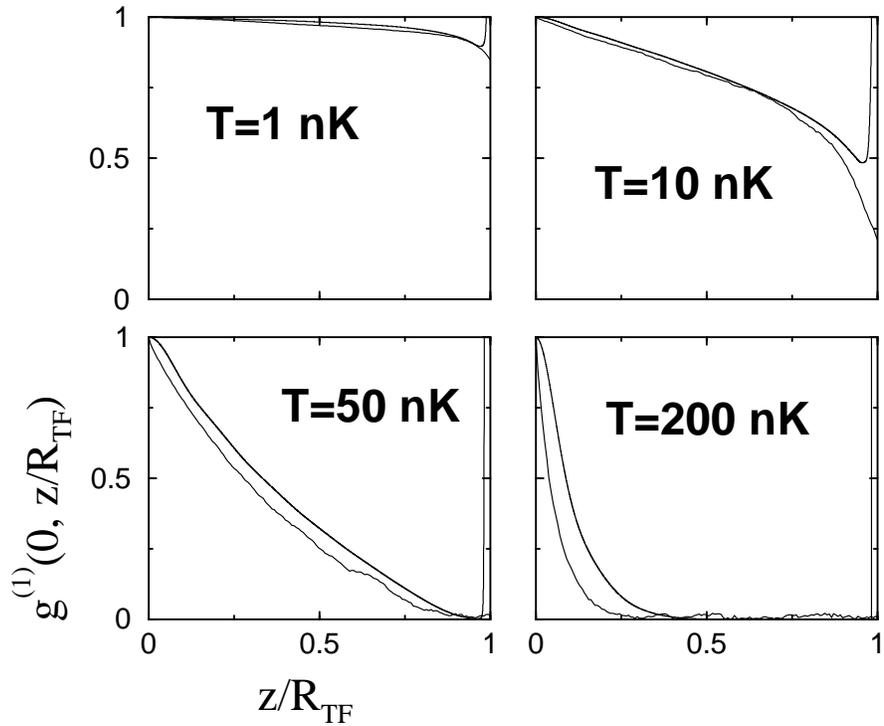}}
\end{center}
\caption[a]{
Comparison of the normalized first-order (phase) correlation functions
calculated using the present mean-field approach, given 
by the solid curves, and the numerical solution of the noisy Langevin 
equation in Eq.~(\ref{lang}) shown with the noisy curves.}
\label{fig10}
\end{figure}

In Figs.~\ref{fig9} and~\ref{fig10}, we show the comparison of the 
many-body $T$-matrix theory
to the above Langevin calculations, for the same temperatures
used in Figs. \ref{fig6} and \ref{fig8}. In Fig.~\ref{fig9} we
compare the Langevin 
densities $\langle|\Phi(z,t)|^{2}\rangle$ to our classical  
mean-field density $n(z)$. This yields 
excellent agreement at low temperatures, except for a small region 
around the discontinuity in the mean-field theory, which can be understood 
from the fact that the local-density approximation always fails in a small 
region near the edge of the Thomas-Fermi radius.
As expected, 
this region increases with
increasing temperature. For $T=50$ nK, Fig. \ref{fig9} further 
shows the deviation of the ``classical'' prediction 
of our mean-field theory from the ``quantum'' one calculated previously 
in Sec.~\ref{prof} and displayed in Fig.~\ref{fig6}. 
Finally, Fig. \ref{fig10} shows the corresponding phase 
correlation functions as a function of position. Here we also find very
good agreement in the entire temperature range. Note that the phase 
correlation functions are essentially indistinguishable for both 
classical and
quantum treatments of the thermal cloud.


It is interesting to note that
the Langevin method yields continuous curves at the expense of 
computational time, due to the large number of independent runs 
that are required
to reduce the statistical error. However,
the Langevin method enables also a direct calculation of the
time-dependent correlation properties via temporal autocorrelation
measurements. Results of such studies, which are of the interest for 
the physics of an atom laser, will be
presented in a separate publication~\cite{nick}.

\begin{figure}[htb]
\epsfysize=9cm
\begin{center}
\centerline{\epsffile{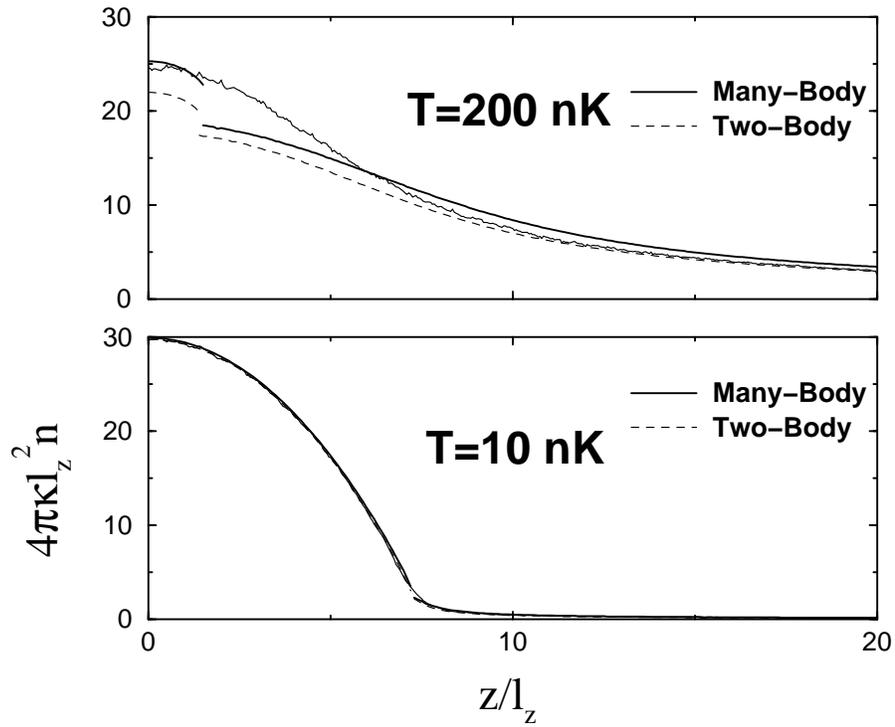}}
\end{center}
\caption[a]{
Study of the many-body renormalization effects 
on the density profiles. The exact results 
are also shown with the noisy curves.
}
\label{fig11}
\end{figure}

\begin{figure}[htb]
\epsfysize=9cm
\begin{center}
\centerline{\epsffile{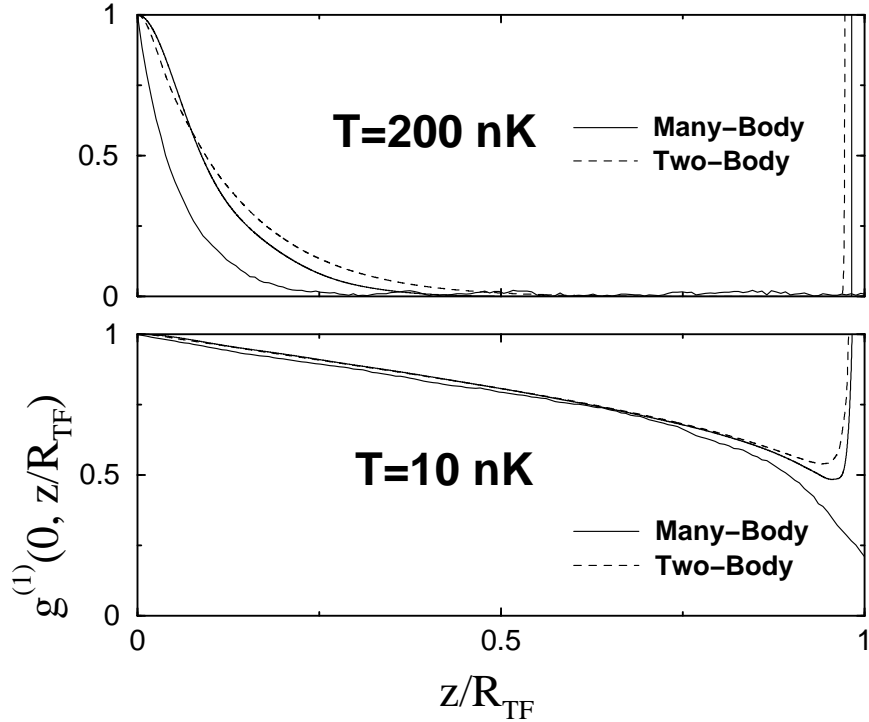}}
\end{center}
\caption[a]{
Study of the many-body renormalization effects 
on the phase correlation function. The exact results 
are also shown with the noisy curves.}
\label{fig12}
\end{figure}

\begin{figure}[htb]
\epsfysize=9cm
\begin{center}
\centerline{\epsffile{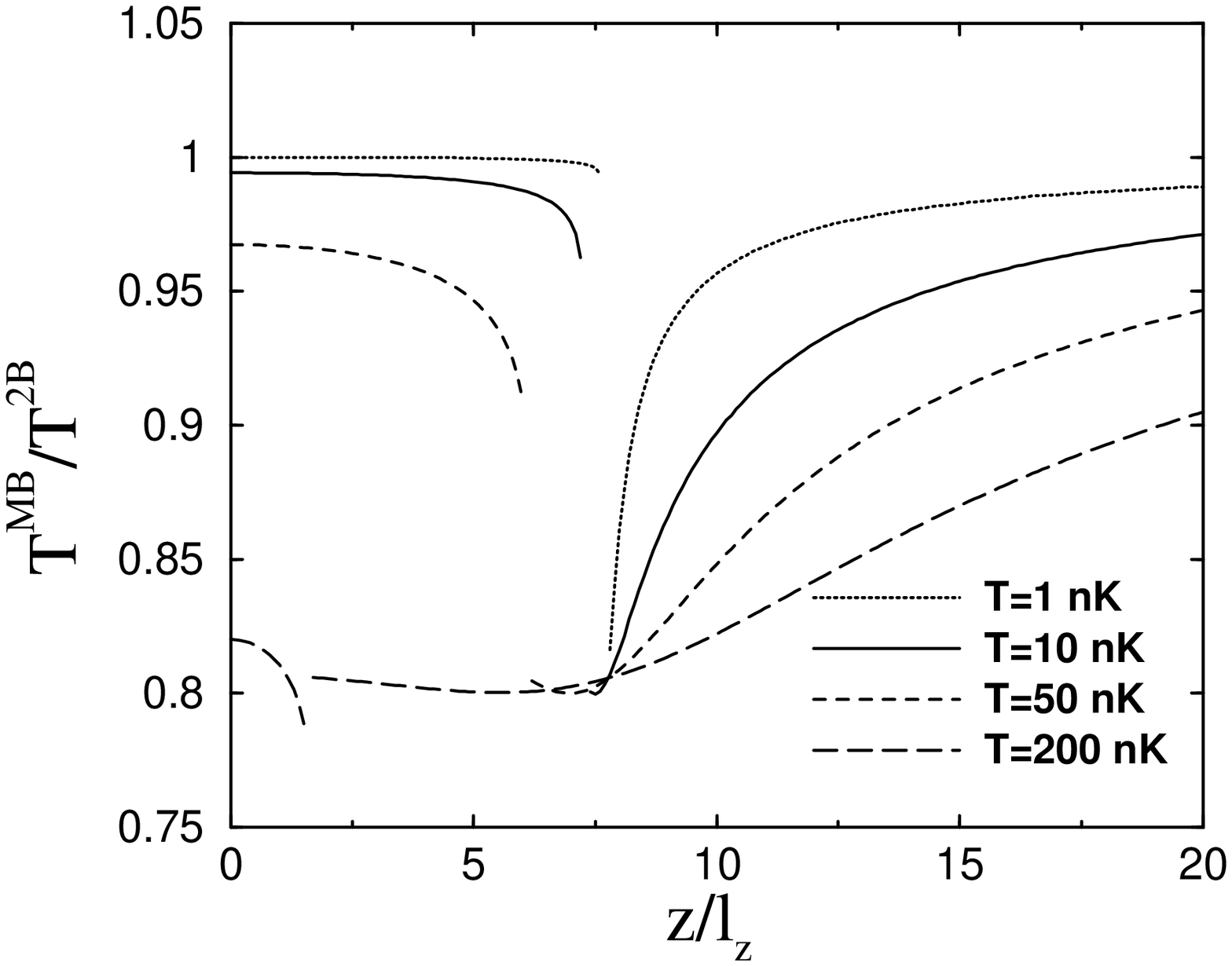}}
\end{center}
\caption[a]{
The many-body T-matrix $T^{\rm MB}$ as a function of the 
distance from the center of the 
trap, for four different temperatures.}
\label{fig13}
\end{figure}

Finally, it is worth mentioning again that 
in obtaining our analytical expressions for the phase fluctuations and the 
density in sections II and V, we have used the 
many-body T-matrix for the interatomic interactions.  
As mentioned in section~\ref{mb} the many-body effects are important in 
one and two dimensions. To appreciate this importance, we recalculate the 
density profiles and phase fluctuations using the two-body T-matrix. 
Thus for 
distances below $R_{\rm TF}$ the differences are due to~Eq. (\ref{mbeq}), 
whereas for distances above $R_{\rm TF}$ they are a result of 
Eq.~(\ref{mbeq2}).
In Figs.~\ref{fig11} and~\ref{fig12} it is clearly seen
that the inclusion of many-body 
effects has led to a better agreement with the exact Langevin results.
Moreover,
the many-body corrections become more pronounced at higher temperatures. 
In Fig.~\ref{fig13}, we show how the renormalized interatomic interaction
strength $T^{\rm MB}(-2\mu(z))$ depends on position. 
We notice that the effects of this renormalization becomes most significant 
near the edge of the condensate and 
for temperatures closer to the 
transition temperature, as expected from the results of 
Refs.~\cite{{rgpaper},{mbpaper}}.


\section{Conclusions}
The Popov theory suffers from infrared divergences 
in the equation of state at all temperatures in one dimension
and at any nonzero temperature in two dimensions. 
These infrared divergences can be 
traced to an inaccurate treatment of the phase fluctuations.
We have proposed a new mean-field theory for dilute Bose gases, 
in which the phase fluctuations are treated exactly.
We have also used this to arrive at an improved
many-body $T$-matrix theory. 
The resulting equation of state is free of infrared divergences
and the theory can thus be applied in any dimension.
Our modified many-body $T$-matrix theory is capable of reproducing exact 
results in one dimension and the
results in three dimensions are to a large extent the same as those predicted
by Popov theory. 
We have used the theory to calculate the reduction of the
recombination rate of a 
spin-polarized two-dimensional hydrogen system. Comparing our calculated rate
with the observed values we found reasonable agreement, although
more work is required to make a detailed comparison.

We have also applied the theory to the Kosterlitz-Thouless phase transition. 
The modified many-body $T$-matrix theory
is used to calculate
initial conditions for the superfluid density and the
fugacity of the vortices in 
a renormalization group calculation that incorporates the physics of
vortex pairs. 
We have calculated the critical temperature for a fixed value of the
$s$-wave scattering length as a function of density, and it was found
that $T_c$ increases almost linearly with density. 
More precisely we have obtained $n\Lambda_c^2\simeq7$.
We believe that this result gives a lower bound on the critical
temperature, since the Kosterlitz-Thouless renormalization group
equations do not include quantum effects, which in principle
affect nonuniversal quantities.

The modified many-body $T$-matrix theory was also applied to calculate 
density profiles 
and phase correlation functions of a one-dimensional trapped Bose gas 
for a variety of temperatures. At very low temperatures,
the phase correlation function was found to decrease only very slightly
over the size of the system, indicating that the equilibrium state
contains a true condensate. 
At larger temperatures, it decreases faster, and the gas now contains 
only a quasicondensate. 
In future work, we will look in detail at this and also at the full
crossover problem between one, two, and three dimensions. Finally,
the densities and phase correlation functions predicted by our 
mean-field theory for various temperatures were compared to the
corresponding predictions of a nonlinear Langevin field equation, 
which gives numerically exact results. The agreement was found to be very 
good for the entire temperature range studied.

\section*{acknowledgments}
We thank Tom Bergeman, Steve Girvin, and Subir Sachdev for valuable
discussions and inspiration. 
We also thank Simo Jaakkola and Sasha Safanov for
providing us with the data of their experiment. 
This work was supported by the Stichting voor 
Fundamenteel Onderzoek der Materie
(FOM), which is supported by the Nederlandse Organisatie voor Wetenschapplijk
Onderzoek (NWO).

\end{multicols}

\newpage

\section*{FIGURES}

\begin{thebibliography}{99}

\bibitem{lowdketterle} A. G\"orlitz, J.M. Vogels, A.E. Leanhardt,
C. Raman, T. L. Gustavson, J. R. Abo-Shaeer, A. P. Chikkatur, S.
Gupta, S. Inouye, T. P. Rosenband, D. E. Pritchard, and W. Ketterle,
Phys. Rev. Lett. {\bf 87}, 130402 (2001).

\bibitem{exp100} F. Schreck, L. Khaykovich, K. L. Corwin, G. Ferrari, 
T. Bourdel, J. Cobizolles, and C. Salomon, Phys. Rev. Lett. {\bf 87}, 
080403 (2001).

\bibitem{exp200} H. Ott, J. Fortagh, G. Schlotterbeck, A. Grossmann, 
and C. Zimmermann, Phys. Rev. Lett. {\bf 87}, 
230401 (2001).

\bibitem{exp300} W. H\"ansel, P. Hommelhoff, T. W. H\"ansch, 
and J. Reichel, Nature {\bf 413}, 501 (2001).

\bibitem{mullin} W. J. Mullin, J. Low. Temp. Phys. {\bf 106}, 615 (1997).

\bibitem{jason} T-L. Ho and M. Ma, J. Low Temp. Phys. {\bf 115}, 61 (1999). 

\bibitem{2d} D. S. Petrov, M. Holzmann, and G.V. Shlyapnikov,
Phys. Rev. Lett. {\bf 84}, 2551 (2000).

\bibitem{1d} D. S. Petrov, G.V. Shlyapnikov, and J.T.M. Walraven,
Phys. Rev. Lett. {\bf 85}, 3745 (2000).

\bibitem{mermin} N. D. Mermin and H. Wagner, Phys. Rev. Lett. {\bf 22}, 1133 
(1966)

\bibitem{hohen} P. C. Hohenberg, Phys. Rev. {\bf 158}, 383 (1967).

\bibitem{popov} V. N. Popov, Theor. Math. Phys. {\bf 11}, 565 (1972);
{\it Functional Integrals in Quantum Field Theory and Statistical
Physics}, (Reidel, Dordrecht, 1983), Chap. 6.

\bibitem{jh}J. O. Andersen, U. Al Khawaja, 
and H. T. C. Stoof, Phys. Rev. Lett {\bf 88}, 070407 (2002). 

\bibitem{koster} J. M. Kosterlitz and D. J. Thouless,
J. Phys. C {\bf 6}, 1181 (1973).

\bibitem{Stoof_Booklet} H. T. C. Stoof, J. Low Temp. Phys. {\bf 114}, 11 (1999).

\bibitem{saf} A. I. Safonov, S.A. Vasilyev, I. S. Yasnikov,
I.I. Lukashevich, and S. Jaakkola, Phys. Rev. Lett. {\bf 81}, 4545 (1998).

\bibitem{henk1} H. T. C. Stoof and M. Bijlsma, Phys. Rev. E {\bf 47}, 939 (1993).

\bibitem{schick} M. Schick, Phys. Rev. A {\bf 3}, 1067 (1971).

\bibitem{fisher} D. S. Fisher and P.C. Hohenberg, Phys. Rev. B
{\bf 37}, 4936 (1988).

\bibitem{rgpaper} M. Bijlsma and H. T. C. Stoof, Phys. Rev. A {\bf 54}, 5085 (1996).

\bibitem{nick1} N. P. Proukakis, S. A. Morgan, S. Choi, and K. Burnett, 
Phys. Rev. A {\bf 58}, 2435 (1998).
\bibitem{snoopy} H. T. C Stoof, M. Bijlsma, and M. Houbiers, J. Res.
Natl. Stand. Technol. {\bf 101}, 443 (1996). 

\bibitem{haldane} F. D. M. Haldane, Phys. Rev. Lett. {\bf 47}, 1840 (1981).

\bibitem{olshanii} M. Olshanii, Phys. Rev. Lett. {\bf 81}, 938 (1998).

\bibitem{GW} M.D. Girardeau and E. M. Wright,
Phys. Rev. Lett. {\bf 84}, 5239 (2000).

\bibitem{henk2}H.T.C. Stoof and M. Bijlsma, Phys. Rev. B {\bf 49}, 422 (1994).

\bibitem{henk3} H. T. C. Stoof, L. P. H. de Goey, W. M. H. M. Rovers,
P. S. M. Kop Jansen, and B.J. Verhaar,  Phys. Rev. A {\bf 38},
1248 (1988).


\bibitem{kagan} Yu. Kagan, B. V. Svistunov, G. V. Shlyapnikov, JETP Lett. {\bf 42}, 
209 (1985).


\bibitem{leeyang} T. D. Lee and C. N. Yang, Phys. Rev. {\bf 105}, 1119 (1957).

\bibitem{girvinprivate} S. M. Girvin, private communication.

\bibitem{koster2} J.M. Kosterlitz, J. Phys. C {\bf 7}, 1046 (1974).

\bibitem{amit} D. J. Amit, Y.Y Goldschmidt, and G. Grinstein, J. Phys. A, 
{\bf 13}, 585 (1980).

\bibitem{minn1}P. Minnhagen and M. Nyl\'en, Phys. Rev. B {\bf 31}, 5768 (1985).

\bibitem{van} W Ketterle and N. J. van Druten, Phys. Rev. A {\bf 54}, 
656 (1996).



\bibitem{fetterusa} A. L. Fetter and D. Rokhsar, Phys. Rev. A {\bf 57}, 
1191 (1998).

\bibitem{stringari} S. Stringari, Phys. Rev. Lett. {\bf 77}, 2360 (1996). 

\bibitem{ghora}D. S. Petrov, G. V. Shlyapnikov, and J. T. M. Walraven, 
Phys. Rev. Lett. {\bf 85}, 3745 (2000).

\bibitem{Stoof_Noisy} H. T. C. Stoof and M.J. Bijlsma, J. Low Temp. Phys. 
{\bf 124}, 431 (2001).


\bibitem{nick}N.P. Proukakis and H.T.C Stoof, unpublished.

\bibitem{mbpaper} M. Bijlsma and H. T. C. Stoof, Phys. Rev. A {\bf 55}, 498
(1997).


\end{thebibliography}
\end{document}